\documentclass[12pt]{article}
\begin{document}
\begin{titlepage}
\begin{flushright}
hep-th/0608143\\
TIT/HEP-558\\
August, 2006\\
\end{flushright}
\vspace{0.5cm}
\begin{center}
{\Large \bf 
Non(anti)commutative $\mathcal{N}=2$ Supersymmetric
Gauge Theory from Superstrings
in the Graviphoton Background
}
\lineskip .75em
\vskip2.5cm
{\large Katsushi Ito and Shin Sasaki}
\vskip 2.5em
{\large\it Department of Physics\\
Tokyo Institute of Technology\\
Tokyo, 152-8551, Japan}  \vskip 4.5em
\end{center}
%\vskip1cm
\begin{abstract}
We study open string amplitudes with 
the D3-branes in type IIB superstring theory compactified on 
${\bf C}^2/{\bf Z}_2$.
We introduce constant graviphoton background along the branes and 
calculate disk amplitudes using the NSR formalism.
We take the zero slope limit and investigate the effective Lagrangian on the D3-branes 
deformed by the graviphoton
background.
We find that the deformed Lagrangian agrees with that of
$\mathcal{N}=2$ supersymmetric $U(N)$ gauge theory
defined in non(anti)commutative $\mathcal{N}=1$ superspace 
 by choosing appropriate graviphoton background. It is also shown that 
abelian gauge theory defined in $\mathcal{N} = 2$ harmonic superspace 
with specific non-singlet deformation is consistent with the deformed  
theory.

\end{abstract}
\end{titlepage}

\baselineskip=0.7cm
\section{Introduction}

It is known that the graviphoton effects play an important
role for studying non-perturbative properties in superstring theory and
supersymmetric gauge theory.
The low energy dynamics of D-branes in the superstrings compactified on
the Calabi-Yau manifold with the constant graviphoton background
 is shown to become supersymmetric gauge theories on non(anti)commutative
superspace \cite{OoVa,BeSe,DeGrNi}.
It is shown that the effective theory becomes 
supersymmetric
Yang-Mills theory on $\mathcal{N}=1/2$ superspace, which was constructed
by Seiberg \cite{Se}.
This theory is defined in $\mathcal{N}=1$ Euclidean superspace 
by introducing non-anticommutativity for supercoordinates
$\theta^\alpha$ satisfying the Clifford algebra $\{
\theta^\alpha,\theta^\beta\}=C^{\alpha\beta}$ \cite{ScNi, KlPeTa}. 
This theory is also considered as the low energy effective theory on 
the D3-branes
of type IIB superstring theory compactified on ${\bf R}^6/{\bf
Z}_2\times {\bf Z}_2$ with constant graviphoton background \cite{Billo-1/2}.

Non(anti)commutative $\mathcal{N}=1/2$ superspace can be
generalized to extended superspace.
Non(anti)commutative harmonic superspace \cite{IvLeZu} 
provides particularly an efficient tool
for investigating the deformed Lagrangian and their symmetries
at the off-shell level.
The $\mathcal{N}=2$ supersymmetric gauge theory on the non(anti)commutative
harmonic superspace has been studied in 
\cite{FeSo,FeIvLeSoZu,Castro,ArItOh2, ArIt3}, 
where  one can
introduce various types of deformation parameters by
$\{ \theta^{i\alpha}, \theta^{j \beta}\}=C^{\alpha\beta ij}$.
Here $\theta^{i\alpha}$ is the supercoordinate labeled by
$SU(2)_R$ $R$-symmetry index $i=1,2$.

The purpose of the present paper is to study the graviphoton effects in
$\mathcal{N}=2$ supersymmetric gauge theory, which can be obtained as the
low-energy
effective theory of the D3-brane in type IIB superstring theory.
We will consider the (fractional) 
D3-branes in type IIB superstring theory compactified on 
the orbifold ${\bf C}^2/{\bf Z}_2$ \cite{Billo-N=2}.
We introduce constant graviphoton backgrounds along the branes and 
calculate disk amplitudes which remain nonzero in the zero slope limit.
Here we will use the NSR formalism to introduce the graviphoton vertex
operator in the closed string R-R sector.
We construct the effective Lagrangian deformed by the graviphoton
background.
The constant graviphoton field strength ${\cal F}^{\alpha\beta ij}$
characterizes the deformation structure of the $\mathcal{N}=2$
supersymmetric gauge theory on the branes.

There arise some non-trivial problems to compare two  parameters
${\cal F}^{\alpha\beta ij}$ and $C^{\alpha\beta ij}$.
One is the choice of the scaling limit 
$(2\pi\alpha')^n{\cal F}=C=\mathrm{fixed}$ for some $n$ in 
the zero slope limit $\alpha'\rightarrow 0$.
Here we take ${\cal F}$ such that it has mass dimension two.
In this work we will fix $n=3/2$ such that $C$ becomes deformation
 parameters of non(anti)commutative superspace. 
Another point is the tensor structure of the graviphoton background.
In the case of superstrings, spinor indices $\alpha,\beta$ and
$R$-symmetry indices $i,j$ are independent.
But in the harmonic superspace formalism the deformation parameter 
$C^{ij}_{\alpha\beta }$ obeys symmetry
$C^{ij}_{\alpha\beta}=C^{ji}_{\beta\alpha}$.
This suggests that the graviphoton background ${\cal F}^{\alpha\beta i j}$ 
describes more general deformation of $\mathcal{N}=2$ theory.
We can classify the graviphoton background into four types
${\cal F}^{[\alpha\beta][ij]}$, ${\cal F}^{(\alpha\beta)[ij]}$, 
${\cal F}^{[\alpha\beta](ij)}$, ${\cal F}^{(\alpha\beta)(ij)}$.
Here the (square) bracket means (anti)symmetrization.

In this paper, we will study the ${\cal F}^{(\alpha\beta)(ij)}$
type deformation in detail.
We will show that in the graviphoton background of type
${\cal F}^{(\alpha\beta)(ij)}$, 
the deformed Lagrangian includes that of
$\mathcal{N}=2$ supersymmetric $U(N)$ gauge theory
defined in  $\mathcal{N}=1/2$ superspace \cite{ArItOh1}.
For the singlet type deformation
$C^{ij}_{\alpha\beta}=C_{s}\epsilon^{ij}\epsilon_{\alpha\beta}$ 
\cite{FeSo}, it is pointed that the deformed
theory can be obtained from the constant R-R scalar background \cite{IvLeZu}.
This deformation would correspond to the ${\cal F}^{[\alpha\beta][ij]}$
type deformation.
However, 
for other types of graviphoton background 
${\cal F}^{[\alpha\beta](ij)}$ and ${\cal F}^{(\alpha\beta)[ij]}$,
they do not correspond to the
deformed theory obtained from the non(anti)commutative harmonic
superspace  due to the 
difference of the tensor structures of indices.

Recently, Bill\'o et. al. \cite{Billo-N=2} studied the low-energy
effective action in
particular type constant graviphoton background and pointed its relation to the
$\Omega$-background which has been applied to obtain the exact prepotential
formula \cite{Ne}.
They use the deformation of type ${\cal F}^{(\alpha\beta)[ij]}$ and
different scaling 
$(2 \pi \alpha')^{\frac{1}{2}} {\cal F}^{(\alpha\beta)[ij]}=\mbox{fixed}$.

This paper is organized as follows:
In section 2, we review type IIB superstrings on ${\bf C}^2/{\bf Z}_2$
using NSR formalism and 
construct
 $\mathcal{N}=2$ supersymmetric
$U(N)$ gauge theory in terms of the fractional D3-branes located at the
singular point in the orbifold
${\bf C}^2/{\bf Z}_2$.
We introduce auxiliary field vertex operators to simplify 
calculations of disk amplitudes.
In section 3, we calculate the disk amplitudes with insertion
of one graviphoton vertex operator.
We focus on the $\mathcal{F}^{(\alpha \beta) 
(ij)}$ type background. In the case that only $\mathcal{F}^{(\alpha 
\beta) (11)}$ is non-zero, the deformed Lagrangian 
is shown to precisely agrees with the 
one that is constructed in $\mathcal{N} = 1$ non(anti)commutative 
superspace. We also show that by restricting to the abelian case, 
the deformed Lagrangian corresponds to the one which is defined in the non-singletly 
deformed harmonic superspace $\{\theta^{i \alpha}, \theta^{j \beta}\} = 
C^{\alpha \beta} b^{ij}$ with $b^{ij} b_{ij} = 0$.
In section 4, we present our conclusions and discuss the possibility of new 
type of deformed $\mathcal{N} = 2$ gauge theory, that is obtained from
the open superstring amplitudes.
In appendix A, we summarize possible one graviphoton disk 
amplitudes which remain nonzero in the zero slope limit. 
In appendix B, we present some details for the effective rules in
computing disk amplitudes including spin operators.

\section{Type IIB superstrings on ${\bf C}^2/{\bf Z}_2$ and 
D3-branes}
In this section we review the construction of the $\mathcal{N}=2$
supersymmetric gauge theory with gauge group $U(N)$ by a stack of
fractional 
D3-branes in type IIB superstring theory compactified on ${\bf C}^2/{\bf
Z}^2$. We will use the NSR formalism.

\subsection{Type IIB on  ${\bf C}^2/{\bf Z}_2$}
We begin with reviewing type II superstring theory in ten dimensions.
Let $X^{m}(z,\bar{z})$, $\psi^m(z)$ and $\tilde{\psi}^m(\bar{z})$ 
($m=1,\cdots,10$)
be free bosons and
fermions with worldsheet coordinates $(z,\bar{z})$.
Here we will take the Euclidean signature and their operator product
expansions (OPEs) are given by
$X^m(z)X^n(w)\sim -\delta^{mn}\ln(z-w)$ and
$\psi^m(z)\psi^n(w)\sim \delta_{mn}/(z-w)$.
Fermionic ghost system $(b,c)$ with conformal weight ($2,-1$) 
and bosonic ghost system $(\beta,\gamma)$ with weight ($3/2,-1/2$) are
also introduced. 
The worldsheet fermions $\psi^m(z)$ are bosonized in terms of free
bosons $\phi^a(z)$ $(a=1,\cdots,5)$ by
\begin{eqnarray}
 f^{\pm e_a}(z)&\equiv&{1\over\sqrt{2}}(\psi^{2a-1}\mp i \psi^{2a})
=: e^{\phi^a}(z): c_{e^a}.
\end{eqnarray}
Here $\phi^a(z)$ satisfy the OPE $\phi^a(z)\phi^b(w)\sim
\delta^{ab}\ln(z-w)$ and the vectors $e_a$ are orthonormal basis 
 in the $SO(10)$ weight lattice space and
$c_{e^a}$ is a cocycle factor \cite{KoLeLeSaWa}.
The bosonic ghost is also bosonized \cite{FMS}:
$
\beta=\partial\xi e^{-\phi}$,
$
\gamma= e^{\phi}\eta
$
with OPE $\phi(z)\phi(w)\sim -\ln(z-w)$. We will omit normal ordering symbol
$: \ :$ sometimes.
In order to describe the R-sector, we need to introduce spin fields
$S^{\lambda}(z)=e^{\lambda\phi}(z)c_{\lambda}$, where $\phi = \phi^a 
e_a$ and $\lambda
={1\over2}(\pm e_1\pm e_2\pm e_3\pm e_4\pm e_5)$.
 $\lambda$ belongs to the spinor representation of $SO(10)$.
$c_{\lambda}$ is a cocycle factor.
In type IIB theory, after the GSO projection, 
we have spinor fields which have
odd number of minus signs in $\lambda$, 
for both left and right movers.

We compactify the theory on ${\bf C}\times {\bf C}^2/{\bf Z}_2$ with
internal coordinates $(x^5,\cdots, x^{10})$ and
put the D3-branes with world volume in $(x^1, x^2,x^3,x^4)$ directions.
We introduce complex string coordinates and worldsheet fermions:
\begin{eqnarray}
 Z&=& {1\over\sqrt{2}} (X^5+i X^6), \quad 
\Psi={1\over\sqrt{2}}(\psi^5+i\psi^6),\nonumber\\
Z^1&=& {1\over\sqrt{2}}(X^7+i X^8), \quad
\Psi^1={1\over\sqrt{2}}(\psi^7+i\psi^8),\nonumber\\
Z^2&=& {1\over\sqrt{2}}(X^9+i X^{10}), \quad
\Psi^2={1\over\sqrt{2}}(\psi^9+i\psi^{10}).
\end{eqnarray}
The ${\bf Z}_2$ action $g$ acts on string coordinates as 
$(Z,Z_1,Z_2)\rightarrow (Z,-Z_1,-Z_2)$.
For spinor states,  $g$ acts as
$+\pi$ rotation on the $7-8$ and $-\pi$ rotation on the $9-10$ plane.
Namely  for a spin state 
$|\lambda_3, \lambda_4,\lambda_5\rangle$,
$g$ acts as $1\otimes i\sigma_3\otimes (-i\sigma_3)$, which breaks the
$SO(6)$ spin symmetry into $SO(2)\times SU(2)$.
${\bf Z}_2$ invariant states are made of
$$
\left|{\epsilon\over2},\pm {1\over2},\pm{1\over2} \right\rangle, \quad 
\epsilon=\pm 1.
$$
Ten-dimensional spinor  field $S^{\lambda}$ 
can be decomposed into $SO(4)\times
SO(2)\times SU(2)$ under the orbifold projection:
\begin{equation}
 S^\lambda \rightarrow (S^{\alpha}S^{(-)}S^{i}, S^{\dot{\alpha}}S^{(+)}S^{i})
\label{eq:decomp1}
\end{equation} 
where
$S^{\alpha}$ and $S^{\dot{\alpha}}$ ($\alpha,\dot{\alpha}=1,2$) spinors are
four-dimensional spinors with weights $\pm {1\over2}(e_1+e_2)$ and
$\pm {1\over2}(e_1-e_2)$, respectively. We will follow the conventions 
of \cite{WeBa}. The upper and lower four-dimensional spinor indices are 
related by the 
anti-symmetric tensor $ \varepsilon^{\alpha \beta}$. $S^{(\pm)}
=e^{\pm{1\over2}\phi_3}$ and
$S^{i}$ denote the internal spin fields. The $S^i$ has weight $\pm{1\over2}(e_4+e_5)$. 
Similarly to the four-dimensional spinors, the internal spin indices $i$
are raised and lowered 
by $\varepsilon^{ij}$.

When $N$ D3-branes are located 
 at the orbifold fixed point, the massless states 
describe $\mathcal{N}=2$ supersymmetric $U(N)$ gauge theory.
The $\mathcal{N}=2$ vector multiplet consists of
gauge bosons $A_\mu$, two gauginos $\Lambda^{\alpha i}$
$(i=1,2)$ and complex scalars $\varphi$, which belong to the adjoint
representation of the gauge group.

We denote the vertex operator for a massless field $X$ in the $q$-picture
 by $V^{(q)}_{X}$.
For bosonic fields in the $\mathcal{N}=2$ vector multiplet, they are given by
\begin{eqnarray}
 V^{(-1)}_{A}&=&(2\pi \alpha')^{1\over2}
A^{\mu}(p) {1\over\sqrt{2}}\psi_\mu e^{-\phi}e^{i\sqrt{2\pi \alpha'}p\cdot X},
 \nonumber \\
 V^{(-1)}_{\varphi}&=&(2\pi \alpha')^{1\over2}
\varphi(p) {1\over\sqrt{2}}\Psi  e^{-\phi}e^{i\sqrt{2\pi \alpha'}p\cdot X},
 \nonumber \\
V^{(-1)}_{\bar{\varphi}}&=&(2\pi \alpha')^{1\over2}
\bar{\varphi} (p)
 {1\over\sqrt{2}}\bar{\Psi}  e^{-\phi}e^{i \sqrt{2\pi \alpha'}p\cdot X},
\end{eqnarray}
where $p^\mu$ is the four-momentum.
For calculations of scattering amplitudes, we need vertex operators in
$0$-picture. These are given by
\begin{eqnarray}
 V^{(0)}_{A}&=&2i (2\pi \alpha')^{1\over2}
A^{\mu}(p) 
\left(\partial X^{\mu}+i (2\pi \alpha')^{1\over2}p\cdot \psi \psi^{\mu}\right)
e^{i\sqrt{2\pi \alpha'}p\cdot X},
\nonumber \\
 V^{(0)}_{\varphi}&=&2i (2\pi \alpha')^{1\over2}
\varphi(p) 
\left(\partial Z+i (2\pi \alpha')^{1\over2}p\cdot \psi \Psi\right)
e^{i\sqrt{2\pi \alpha'}p\cdot X},
\nonumber \\
V^{(0)}_{\bar{\varphi}}&=&2i (2\pi \alpha')^{1\over2}
\bar{\varphi}(p) 
\left(\partial \bar{Z}+i (2\pi \alpha')^{1\over2}p\cdot \psi \bar{\Psi}\right)
e^{i\sqrt{2\pi \alpha'}p\cdot X}.
\end{eqnarray}
For fermionic fields, they are constructed by using the spin fields:
\begin{eqnarray}
 V^{(-1/2)}_{\Lambda}&=& 
(2\pi \alpha')^{3\over4}
\Lambda^{\alpha i}(p)S_\alpha S^{(-)}S_i 
e^{-{1\over2}\phi}e^{i\sqrt{2\pi \alpha'}p\cdot X}, \nonumber \\
 V^{(-1/2)}_{\bar{\Lambda}}&=& 
(2\pi \alpha')^{3\over4}
\bar{\Lambda}_{\dot{\alpha} i}(p)\bar{S}^{\dot{\alpha}} 
S^{(+)}S^i 
e^{-{1\over2}\phi}e^{i\sqrt{2\pi \alpha'}p\cdot X}.
\end{eqnarray}
The prefactor of the vertex operators ensures 
that all the polarization has canonical dimension. Following \cite{Billo-1/2}, 
the Fourier transformation is taken with respect to the dimensionless 
momentum $k \equiv \sqrt{2 \pi \alpha'} p$ so that the momentum polarization 
$A_{\mu}(p)$ has the same dimension of $A_{\mu} (x)$. 

The graviphoton vertex operator belongs to the R-R sector and is 
expressed as
\begin{eqnarray}
 V^{(-1/2,-1/2)}_{{\cal F}}(z,\bar{z})&=&
(2\pi\alpha') {\cal F}^{\alpha\beta ij}
e^{-{1\over2}\phi}S_\alpha S^{(-)}S_{i}(z)
e^{-{1\over2}\phi}\tilde{S}_\beta \tilde{S}^{(-)}\tilde{S}_{j}(\bar{z}).
\end{eqnarray}
We normalized  $\mathcal{F}^{\alpha 
\beta i j}$ such that it has canonical mass dimension $+2$. 

\subsection{Disk amplitudes}
We now consider a disk amplitude such that open strings end on the
D3-branes.
The disk is 
realized as the upper half-plane whose boundary is
real axis.
The vertex operators for massless vector multiplets are inserted on the
real axis and the graviphoton operators are in the upper-half plane.
We apply the doubling trick where right-moving fields are located 
on the lower-half plane with the boundary
condition:
\begin{equation}
\left. S_{\alpha}S^{(-)}S_{i}(z)=\tilde{S}_{\alpha}\tilde{S}^{(-)}
\tilde{S}_{i}(\bar{z})
\right|_{z=\bar{z}} .
\end{equation}
The disk amplitudes can be calculated by replacing
$\tilde{S}_{\alpha}\tilde{S}^{(-)}
\tilde{S}_{i}(\bar{z})$ by
$S_{\alpha}S^{(-)}S_{i}(\bar{z})$ in the correlator.
The $n+2 n_{\cal F}$-point disk amplitude for $n$ vertex operators
$V^{(q_i)}_{X_i}(y_i)$ and $n_{\cal F}$ graviphoton vertex operators
$V^{(-{1\over2},-{1\over2})}_{\cal F}(z_j,\bar{z}_j)$
is given by
\begin{equation}
 \langle \! \langle V^{(q_1)}_{X_1}\cdots V^{(-{1\over2},-{1\over2})}_{\cal
  F}
\cdots \rangle \! \rangle
=C_{D_2}\int {\prod_{i=1}^{n}dy_i \prod_{j=1}^{n_{\cal F}}
dz_jd\bar{z}_j
\over dV_{CKG}}
\langle V^{(q_1)}_{X_1}(y_1)\cdots V^{(-{1\over2},-{1\over2})}_{\cal F}
(z_1,\bar{z_1})\cdots \rangle.
\label{eq:disk1}
\end{equation}
Here $C_{D_2}$ denotes the disk normalization factor, which is given by
\cite{VMLRM}
\begin{equation}
 C_{D_2}={1\over 2\pi^2(\alpha')^2 }{1\over k g_{\mathrm{YM}}^2}
\end{equation}
where $g_{\mathrm{YM}}$ is the gauge coupling constant and $k$ is a normalization of $U(N)$ generators
$T^a$, ${\rm tr}(T^a T^b)=k\delta^{ab}$.
$dV_{CKG}$ is an $SL(2,{\bf R})$-invariant volume factor to fix
three positions $x_1$, $x_2$ and $x_3$ among $y_i$, $z_j$,and
$\bar{z}_j$'s:
\begin{equation}
 d V_{CKG}={d x_1 d x_2 d x_3\over
(x_1-x_2) (x_2-x_3) (x_3-x_1)}.
\end{equation}
Note that in the disk amplitudes (\ref{eq:disk1}) the sum of the 
$\phi$-charge in the bosonic ghost must be $-2$.

We need some correlation functions of ten-dimensional spin operators
including bosonized ghosts such as
$e^{-{1\over2}\phi}S^{\lambda}(z)$, bosonized fermions $f^{\pm
e_i}(z)$ and Lorentz generators $: f^{\pm e_i} f^{\pm e_j} (z):$.
Lorentz generators can be eliminated from the correlation functions by
using the Ward identities (see Appendix B). 
The correlation functions are reduced to the ones of bosonized vertex
operators of the form $e^{\tilde{\lambda}
\cdot\tilde{\phi}}(z)c_{\tilde{\lambda}}
=e^{\lambda\cdot\phi}e^{q\phi}(z)c_{\tilde{\lambda}}$.
Here $\tilde{\lambda}=(\lambda,q)$ and $\tilde{\phi}=(\phi^a,\phi)$.
The cocycle factor is given by
$c_{\tilde{\lambda}}=\exp(\pi i \tilde{\lambda} M  [\partial\tilde{\phi}]_0)$, where
$[\partial\tilde{\phi}]_0$ denotes the zero mode of
$\partial\tilde{\phi}$
and the $6\times 6$ matrix $M$ \cite{KoLeLeSaWa} is given by
\begin{equation}
 M=\left(
\begin{array}{cccccc}
0 & 0 & 0& 0 & 0& 0\\
1 & 0& 0 & 0 & 0& 0\\
1 & 1& 0 & 0 & 0& 0\\
-1 & 1 & -1 & 0 & 0& 0\\
1 & 1 & 1& 1 & 0& 0\\
-1 & -1 & -1& -1 & 1& 0\\
\end{array}
\right).
\label{eq:cocymat1}
\end{equation}
Then the correlation functions are calculated as
\begin{equation}
 \langle
  e^{\tilde{\lambda}_1\cdot\tilde{\phi}}(z_1)c_{\tilde{\lambda}_1}
\cdots e^{\tilde{\lambda}_N\cdot \tilde{\phi}}(z_N)c_{\tilde{\lambda}_2}
\rangle
=\prod_{i<j}(z_i-z_j)^{\tilde{\lambda}_i \tilde{\lambda}_j}
\exp (\pi i \tilde{\lambda}_i \cdot M\tilde{\lambda}_j)
\delta_{\sum_{i}\tilde{\lambda}_i, (0,-2)}.
\end{equation}
Here $\tilde{\lambda}_i\cdot \tilde{\lambda}_j=\lambda_i\cdot
\lambda_j-q_i q_j$ for $\tilde{\lambda}_i=(\lambda_i,q_i)$.
When we decompose the spin operators as in (\ref{eq:decomp1}), we can
obtain the ``effective'' rules for space-time and internal
parts \cite{DIS}. These rules are summarized in Appendix B.

\subsection{$\mathcal{N}=2$ gauge theory and the auxiliary field method}

The action of $\mathcal{N}=2$ supersymmetric Yang-Mills theory is
given by
\begin{eqnarray}
 S^{\mathcal{N}=2}_{\mathrm{SYM}} &=&\int \! d^4 x \
{1\over g_{\mathrm{YM}}^2}  {1\over k}
{\rm tr}
\left(
-{1\over4}F_{\mu\nu}F^{\mu\nu}-{1\over4}F_{\mu\nu}\tilde{F}^{\mu\nu}
-D_{\mu}\varphi D^{\mu}\bar{\varphi}
-{1\over2}\left[\varphi,\bar{\varphi}\right]^2
\right.
\nonumber\\
&&\left.
-i \Lambda^i \sigma^\mu D_\mu \bar{\Lambda}_i
-{1\over\sqrt{2}}\Lambda^i \left[\bar{\varphi},\Lambda_i\right]
-{1\over\sqrt{2}}\bar{\Lambda}_i \left[\varphi,\bar{\Lambda}^i\right]
\right),
\label{eq:lag0}
\end{eqnarray}
where
\begin{eqnarray}
 F_{\mu\nu}&=& \partial_\mu A_\nu-\partial_\nu A_\mu+i [A_\mu, A_\nu], 
  \nonumber \\
D_{\mu}\varphi&=&\partial_\mu \varphi+ i[A_\mu,\varphi],
\end{eqnarray}
and $\tilde{F}_{\mu\nu}$ is the dual of $F_{\mu\nu}$.
$\sigma_\mu=(i\tau^1,i\tau^2,i\tau^3,1)$ and
$\bar{\sigma}_\mu=(-i\tau^1,-i\tau^2,-i\tau^3,1)$ are Dirac matrices.
Here $\tau^a$ ($a=1,2,3$) denote the Pauli matrices.
The gauge fields $A_\mu$ and other fields are expanded such as 
$A_\mu=A_{\mu}^a T^a$.
%, where
%the generators $T^a$ ($a=1,\cdots, N^2$) of the Lie algebra of $U(N)$
%are normalized as ${\rm tr}(T^a T^b)=k \delta^{ab}$.
In the action (\ref{eq:lag0}) 
we have eliminated auxiliary fields of the superfields.
The action is derived by computing disk amplitudes with vertex operators
attached on the boundary of the disk.

The auxiliary field method \cite{Billo-1/2} (see also \cite{DIS,ADS}) is 
found to give an  effective tool to simplify calculations  because 
four-point amplitudes can
 be reduced to an three-point amplitudes which include
an auxiliary field vertex operator.
In \cite{Billo-1/2}, this method was applied to obtain non(anti)commutative
$\mathcal{N}=1/2$ super Yang-Mills theory from the D3-brane in type
IIB superstrings compactified on ${\bf C}^3/{\bf Z}_2\times {\bf Z}_2$.
In this paper we generalize this method to the case of the 
$\mathcal{N}=2$ gauge theory.

In \cite{Billo-1/2}, it was shown that
the quartic interactions of gauge fields can be written into the
cubic type interactions by introducing the auxiliary self-dual
tensor $H_{\mu\nu}$, which is also expressed in terms of 't Hooft 
eta symbol such as $H_{\mu\nu}=H^c \eta^c_{\mu\nu}$.
The gauge field part $-{1\over 4 g^2_{\mathrm{YM}}k} {\rm
tr}(F^{2}_{\mu\nu}+F_{\mu\nu}\tilde{F}^{\mu\nu})$ 
in the Lagrangian is equivalent to
\begin{equation}
-{1\over g_{\mathrm{YM}}^2}{1\over k}{\rm tr}
\left(
{1\over4}(\partial_\mu A_\nu-\partial_\nu A_\mu)^2
+i \partial_\mu A_\nu [A^\mu, A^\nu]
+{1\over2}H_c H^c+{1\over2}H_c \eta^c_{\mu\nu} [A^\mu, A^\nu]
\right).
\label{eq:lag1}
\end{equation}
In the $\mathcal{N}=2$ case,
the action (\ref{eq:lag0}) contains other quartic interactions which
include scalar fields and gauge fields.
We therefore introduce  new auxiliary 
fields  ${H_{A \varphi}}_{\mu}$, ${H_{A \bar{\varphi}}}_{\mu}$ and 
$H_{\varphi\bar{\varphi}}$.
The Lagrangian is shown to be equal to
\begin{eqnarray}
{\cal L}^{\mathcal{N}=2}_{\mathrm{SYM}}&=&
-{1\over g_{YM}^2}{1\over k}{\rm tr}
\left[
{1\over4}(\partial_\mu A_\nu-\partial_\nu A_\mu)^2
+i \partial_\mu A_\nu [A^\mu, A^\nu]
+{1\over2}H_c H^c+{1\over2}H_c \eta^c_{\mu\nu} [A^\mu, A^\nu] \right.
\nonumber\\
&&
+\partial_\mu\varphi\partial^\mu \bar{\varphi}
+i\partial_\mu \varphi [A^\mu, \bar{\varphi}]
+i [A_\mu,\varphi]\partial \bar{\varphi}
-H_{A\varphi \mu}H^{\mu}_{A\bar{\varphi}}
+iH_{A\varphi \mu}[A^\mu, \bar{\varphi}]
+i [A_\mu,\varphi] H^{\mu}_{A\bar{\varphi}}
\nonumber\\
&&+H_{\varphi\bar{\varphi}}^2+i\sqrt{2}H_{\varphi\bar{\varphi}} [\varphi,
\bar{\varphi}] 
\nonumber\\
&& \left.
-i \Lambda^i \sigma^\mu D_\mu \bar{\Lambda}_i
-{1\over\sqrt{2}}\Lambda^i \left[\bar{\varphi},\Lambda_i\right]
-{1\over\sqrt{2}}\bar{\Lambda}_i \left[\varphi,\bar{\Lambda}^i\right]
\right].
\label{eq:lag3}
\end{eqnarray}

The auxiliary fields have relevant vertex operators in superstring theory.
In \cite{Billo-1/2}, it is shown that the auxiliary fields
$H_{\mu\nu}$ is associated to the vertex operator
\begin{equation}
 V_{H}^{(0)}(y)={1\over2}(2\pi \alpha') H_{\mu\nu}(p)\psi^\nu\psi^\mu
e^{i\sqrt{2\pi\alpha'}p\cdot X}(y)
\end{equation}
in the 0-picture.
In the $\mathcal{N}=2$ case this vertex operator can be generalized to other 
auxiliary fields such as
\begin{eqnarray}
 V^{(0)}_{H_{A\varphi}}&=& 2i (2\pi\alpha') {H_{A\varphi}}_{\mu}
\psi^\mu \Psi e^{i\sqrt{2\pi\alpha'}p\cdot X}, \nonumber \\
V^{(0)}_{H_{A\bar{\varphi}}}&=&
2i (2\pi\alpha') {H_{A \bar{\varphi}}}_{\mu}
\psi^\mu \bar{\Psi} e^{i\sqrt{2\pi\alpha'}p\cdot X},\nonumber \\
V^{(0)}_{H_{\varphi\bar{\varphi}}}&=&
-i\sqrt{2}(2\pi\alpha') H_{\varphi \bar{\varphi}}
\Psi\bar{\Psi}e^{i\sqrt{2\pi\alpha'}p\cdot X}.
\end{eqnarray}

We now explain that all the interaction terms in the $\mathcal{N}=2$ 
Lagrangian (\ref{eq:lag3}) 
can be derived from the disk amplitudes with vertex operators on the
boundary. For example, the $H_{\mu \nu} [ A^{\mu}, 
 A^{\nu}]$ term in (\ref{eq:lag3}) is derived from the disk amplitude
\begin{eqnarray}
& & \langle \! \langle V^{(0)}_{H} (p_1) V^{(-1)}_{A} (p_2) 
 V^{(-1)}_{A} (p_3) \rangle \! \rangle \nonumber \\
& & \qquad = \frac{1}{2 \pi^2 \alpha'^2} 
\frac{1}{k g^2_{\mathrm{YM}}}
 (2 \pi \alpha')^{2} \frac{1}{2} \left(\frac{1}{\sqrt{2}} \right)^2 
\mathrm{tr} \left[ H_{\mu \nu} (p_1) 
 A_{\rho} (p_2) A_{\sigma} (p_3) \right]  \int \!\! \frac{\prod_j d y_j}{d 
 V_{\mathrm{CKG}}} \nonumber \\
& & \qquad \qquad \times  \langle e^{- \phi (y_2)} e^{- \phi (y_3) } \rangle 
\langle \psi^{\nu} \psi^{\mu} (y_1) \psi^{\rho} (y_2) 
 \psi^{\sigma} (y_3) \rangle \left\langle \prod_{j=1}^3 e^{i \sqrt{2 \pi 
			      \alpha'} p_j \cdot X (y_j)} \right\rangle. 
\nonumber \\
\end{eqnarray}
Here we have separated the correlator into the four-dimensional, internal 
and ghost parts. The ghost part can be evaluated 
by the Wick formula.
%\begin{eqnarray}
%\langle e^{- \phi (y_2)} e^{- \phi (y_3) } \rangle 
%= (y_2 - y_3)^{-1}.
%\end{eqnarray}
The other parts are calculated by the effective rules in Appendix B.
%The four-dimensional four-fermion correlator is
%\begin{eqnarray}
%\langle \psi^{\nu} \psi^{\mu} (y_1) \psi^{\rho} (y_2) 
% \psi^{\sigma} (y_3) \rangle
%= - \frac{\delta^{\nu \rho} \delta^{\mu \sigma} - \delta^{\nu \sigma} 
%\delta^{\mu \rho}}{(y_1 - y_2) (y_1 - y_3)}.
%\end{eqnarray}
The $X^{\mu}$ correlator is given by 
\begin{eqnarray}
\left\langle \prod_{j=1}^3 e^{i \sqrt{2 \pi 
 \alpha'} p_j \cdot X (y_j)} \right\rangle = 
(y_1 - y_2)^{2 \pi \alpha' p_1 \cdot p_2} 
(y_1 - y_3)^{2 \pi \alpha' p_1 \cdot p_3}
(y_2 - y_3)^{2 \pi \alpha' p_2 \cdot p_3}.
\end{eqnarray}  
Since we consider only the massless states,
we have $p_i \cdot p_j = 0$ for $i,j = 1 \cdots 3$ and the contribution from the $X$ correlator 
becomes trivial.
Taking all together, the amplitude is
\begin{eqnarray}
\langle \! \langle V^{(0)}_{H} (p_1) V^{(-1)}_{A} (p_2) 
 V^{(-1)}_{A} (p_3) \rangle \! \rangle = 
\frac{1}{ g^2_{\mathrm{YM}} k} \mathrm{tr} 
\left[
H_{\mu \nu} (p_1) A^{\mu} (p_2) A^{\nu} (p_3) 
\right].
\end{eqnarray}
We note that the appropriate $\alpha'$ scaling appeared. 
After adding the other inequivalent color ordered amplitudes to the 
above one and taking the symmetric factor into account, 
we find the interaction term corresponding to the amplitude is
\begin{eqnarray}
\mathcal{L} = - \frac{1}{2 g^2_{\mathrm{YM}}} \frac{1}{k} \mathrm{tr}
\left[ H_{\mu \nu} (x) [A^{\mu} (x), A^{\nu} (x)] \right],
\end{eqnarray}
which is precisely the desired interaction in (\ref{eq:lag3}).
The other interaction can be calculated by the same way and the results are
\begin{eqnarray}
& & \left\langle \! \left\langle  V^{(0)}_A (p_1) V^{(-1)}_A (p_2) V^{(-1)}_A 
(p_3) \right\rangle \! \right\rangle  = 
 - \frac{2}{k g^2_{\mathrm{YM}}}
 \mathrm{tr} \left[ A_{\mu} (p_1) p_2^{\mu} A_{\rho} (p_2) A_{\sigma} 
 (p_3) \delta^{\rho \sigma} \right. \nonumber \\
& & \qquad \qquad \qquad \qquad
\qquad \qquad \qquad \qquad \qquad \quad + p_{1 \nu} A_{\mu} (p_1) A_{\rho} (p_2) A_{\sigma} (p_3) \delta^{\mu\rho}
\delta^{\nu \sigma} \nonumber \\
& & \qquad \qquad \qquad \qquad
\qquad \qquad \qquad \qquad \qquad \quad  \left.  - p_{1 \nu} A_{\mu} (p_1) A_{\rho} (p_2) A_{\sigma} (p_3) \delta^{\mu\sigma}
\delta^{\nu \rho} \right], \\
& & \left\langle \! \left\langle  V^{(0)}_{H_{\varphi \bar{\varphi}}} (p_1)
 V^{(-1)}_{\varphi} (p_2) V^{(-1)}_{\bar{\varphi}}(p_3) \right\rangle \! \right\rangle 
=  \frac{i \sqrt{2}}{k g^2_{\mathrm{YM}}} \mathrm{tr}
\left[ H_{\varphi \bar{\varphi}} (p_1) \varphi (p_2) \bar{\varphi} (p_3) 
\right], \\
& & \left\langle \! \left\langle V^{(0)}_{H_{A \varphi}} (p_1) 
V^{(-1)}_{A} (p_2) V^{(-1)}_{\bar{\varphi}}(p_3) \right\rangle \! \right\rangle 
= \frac{2 i}{k g^2_{\mathrm{YM}}} \mathrm{tr} \left[
{H_{A \varphi}}_{\mu} (p_1) A^{\mu} (p_2) \bar{\varphi} (p_3) \right], 
 \\
& & \left\langle \! \left\langle V^{(0)}_{\varphi} (p_1) V^{(-1)}_{A} (p_2) V^{(-1)}_{\bar{\varphi}}(p_3) 
\right\rangle \! \right\rangle
= - \frac{2}{k g^2_{\mathrm{YM}}} \mathrm{tr} \left[
p_{1 \mu} \varphi (p_1) A^{\mu} (p_2) \bar{\varphi} (p_3) \right], \\
& & \left\langle \! \left\langle V^{(-1/2)}_{\Lambda} (p_1) V^{(-1/2)}_{\Lambda} (p_2) 
 V^{(0)}_{\bar{\varphi}} (p_3) \right\rangle \! \right\rangle =  - 
 \frac{\sqrt{2}}{k g^2_{\mathrm{YM}}} \mathrm{tr} \left[ \Lambda^{\alpha i} (p_1) \Lambda_{\alpha i} (p_2) 
 \bar{\varphi} (p_3) \right], \\
& & \left\langle \!\left\langle V^{(-1/2)}_{\Lambda} (p_1) V^{(-1)}_{A} (p_2) 
V^{(-1/2)}_{\overline{\Lambda}} (p_3) \right\rangle \!\right\rangle =
 \frac{1}{k g^2_{\mathrm{YM}}} \mathrm{tr} \left[
\Lambda^{\alpha i} (p_1) (\sigma^{\mu})_{\alpha}^{\ \dot{\beta}} A_{\mu} 
(p_2) \overline{\Lambda}_{\dot{\beta} j} (p_3)
\right].
\end{eqnarray}
Adding other inequivalent color ordered amplitudes and the phase shift 
of $\Lambda$, we find that all the cubic interactions in (\ref{eq:lag3}) 
are reproduced from these disk amplitudes.

\section{Disk amplitudes in the constant graviphoton background}
In this section, we will calculate the correction to the disk amplitudes 
due to the insertion of one graviphoton vertex operator.

\subsection{The zero slope limit}
We now examine the effect of the graviphoton vertex operator
inserted in the disk. We will take the zero slope (field theory)
limit $\alpha'\rightarrow 0$ at the final stage of the amplitudes calculation. 
The R-R graviphoton vertex operator in the disk amplitudes is written as
\begin{eqnarray}
V_{\mathcal{F}}^{(-1/2,-1/2)} (z, \bar{z}) \!\! 
 = \!\! (2 \pi \alpha') \mathcal{F}^{\alpha \beta i j} 
 \left[ S_{\alpha} (z) S^{(-)} (z) S_i (z) e^{- \frac{1}{2} \phi (z)} 
  S_{\beta} (\bar{z}) S^{(-)} (\bar{z}) S_j (\bar{z}) e^{ - \frac{1}{2} 
  \phi (\bar{z})} \right],
\end{eqnarray}
where we identify the left- and right-moving part.

We need to fix the scaling of the constant 
graviphoton background.
In general we can take the limit such that
\begin{equation}
 (2\pi\alpha')^n {\cal F}^{\alpha\beta i j}=C^{\alpha\beta i j}
\label{eq:scaling1}
\end{equation}
is fixed for some $n$.
For $n=3/2$, the parameter $C^{\alpha\beta i j}$ has mass dimensions
$-1$, which has the same dimension as the deformation parameters in
 the non(anti)commutative field theory.
%In the following we will focus on this type of scaling.

We firstly explore which type of disk amplitudes remain nonzero in the zero
slope limit.
Let $n_{X}$ be the number of vertex operators for a massless field $X$.
Assuming that we can assign the appropriate picture number for each
vertex operators, the amplitudes of the form
$\langle \! \langle V_{A}^{n_A}\cdots V^{n_{\bar{\varphi}}}_{\bar{\varphi}}
V_{\cal F}^{n_{\cal F}}\rangle \! \rangle$
scale as $(\alpha')^M$, where
\begin{equation}
M=-2 +{1\over2}(n_A+n_{\varphi}+n_{\bar{\varphi}})
+{3\over4}(n_{\Lambda}+n_{\bar{\Lambda}})+\left(1-n\right)
n_{\cal F}.
\label{eq:scale2}
\end{equation}
Here $-2$ comes from the normalization of the disk amplitudes.
For $M\leq 0$, the amplitudes remains nonzero in the zero-slope limit.
Using the $\phi_3$ charge conservation we get
\begin{eqnarray}
-n_{\varphi}+n_{\bar{\varphi}}-{1\over2}n_{\Lambda}+{1\over2}n_{\bar{\Lambda}}
-n_{\cal F}=0 .
\label{eq:phi3charge}
\end{eqnarray}
Using (\ref{eq:scale2}) and (\ref{eq:phi3charge}), the condition $M \le 
0$ becomes
\begin{equation}
{1\over2}n_{A}+ \left( n-\frac{1}{2} \right) n_{\varphi}+
  \left(\frac{3}{2} - n \right) n_{\bar{\varphi}}
+\frac{1}{2} \left( n + \frac{1}{2} \right) n_{\Lambda}
+ \frac{1}{2} \left( \frac{5}{2} - n \right) n_{\bar{\Lambda}}
\leq 2 .
\label{eq:scale3}
\end{equation}
We then can classify which type of amplitudes remain non-zero in the
zero-slope limit.
This analysis
can be generalized to the amplitudes including auxiliary field
vertex operators.
Let $n^{H}_{Y}$ be 
the number of vertex operators for auxiliary fields $H_{Y}$.
Then the condition (\ref{eq:scale3}) becomes
\begin{equation}
{1\over2}n_{A}+ \left( n - \frac{1}{2} \right) n_{\varphi}+
\left( \frac{3}{2} - n \right) n_{\bar{\varphi}}
+{1+2n\over4}n_{\Lambda}+{5-2n\over4} n_{\bar{\Lambda}}
+n^{H}_{AA}+n^{H}_{\varphi\bar{\varphi}}
+ n n^{H}_{A\varphi}
+(2-n)n^{H}_{A\bar{\varphi}}
\leq 2
\label{eq:scale4}
\end{equation}
with the $\phi_3$-charge conservation
\begin{eqnarray}
-n_{\varphi}+n_{\bar{\varphi}}
-{1\over2}n_{\Lambda}+{1\over2}n_{\bar{\Lambda}}
-n^{H}_{A\varphi}+n^{H}_{A\bar{\varphi}}
-n_{\cal F}
=0.
\label{eq:phi3charge2}
\end{eqnarray}

We now consider the case $n=3/2$. 
In this case, the condition (\ref{eq:scale4}) becomes
\begin{equation}
{1\over2}n_{A}+ n_{\bar{\varphi}}
+n_{\Lambda}+{1\over2} n_{\bar{\Lambda}}
+n^{H}_{AA}+n^{H}_{\varphi\bar{\varphi}}
+{3\over2}n^{H}_{A\varphi}
+{1\over2}n^{H}_{A\bar{\varphi}}
\leq 2.
\label{eq:scale4a}
\end{equation}
Without auxiliary fields, we find that 17 types amplitudes remain
non-zero. For example,
$A^4 \bar{\varphi}^{n_{\cal F}}{\cal F}^{n_{\cal F}}$
type amplitudes remain non-vanishing in the zero slope limit for $n_{\cal F}\geq 0$.
This infinite series type correction 
arises also in the case of non(anti)commutative
harmonic superspace \cite{IvLeZu, ArItOh2}. Indeed, this systematic 
analysis of $\alpha'$ scaling is only a sufficient condition for the 
non-vanishing amplitudes in the field theory limit. In the string theory 
 there is no guarantee that the amplitude is non-vanishing even 
though it has an appropriate $\alpha'$ scaling.

In this work we will consider the lowest order correction to the amplitude
 by one constant graviphoton vertex operator.

\subsection{Disk amplitudes in the zero slope limit
with fixed $(2\pi \alpha')^{\frac{3}{2}}{\cal F}$}
We will examine possible structure of string amplitudes 
in the scaling limit with fixed $(2\pi \alpha')^{\frac{3}{2}}{\cal F}$.
But it is necessary to see the explicit from of the 
correlator before the zero-slope limit is taken.

Focusing on the $\phi_3$ charge, the graviphoton vertex operator contains two 
 internal spin fields $S^{(-)}$. To cancel this $\phi_3$ charge, 
one should insert one $\bar{\varphi}$ vertex or two $\overline{\Lambda} 
$ vertex operators. Thus the non-zero disk amplitudes that include
 one graviphoton vertex operators should be of the form
\begin{eqnarray}
\langle \! \langle \cdots V_{\bar{\varphi}} V_{\mathcal{F}} \rangle \! \rangle,
 \qquad \langle \! \langle \cdots V_{\overline{\Lambda}}  V_{\overline{\Lambda}} 
V_{\mathcal{F}} \rangle \! \rangle,
\label{non-zero_combination}
\end{eqnarray}
where remaining part is $\phi_3$ neutral. Possible insertions are of the form
\begin{eqnarray}
V_A, \ V_{\varphi} V_{\bar{\varphi}}, \ V_{\Lambda} 
 V_{\overline{\Lambda}}, \ V_{\varphi} V_{\overline{\Lambda}} 
V_{\overline{\Lambda}}, \ V_{\bar{\varphi}} V_{\Lambda} V_{\Lambda}.
\label{neutral}
\end{eqnarray} 
Thus, the non-zero disk amplitudes for the one graviphoton vertex insertion have the 
structure of (\ref{non-zero_combination}) with the insertion of the vertex 
operators in (\ref{neutral}). 

As mentioned in \cite{Billo-quiver}, when we have non-zero 
amplitude with a 0-picture vertex operator $V^{(0)}_X$ corresponding to 
the fields $X = (A_{\mu}, \varphi, \bar{\varphi})$ (which produces the derivative $\partial_{\mu} X$), 
 the amplitude in which $V^{(0)}_X$ is replaced by $V^{(0)}_{H_{AX}}$ is 
 also non-zero. The combined amplitude $\langle \! \langle (V^{(0)}_X + V^{(0)}_{H_{AX}}) \cdots 
\rangle \! \rangle $ correspond to the gauge covariant derivative 
$D_{\mu} X$.
Thus, whenever we have non-zero amplitude in the zero-slope limit which contains 
 0-ghost picture vertex operator $V^{(0)}_{X}$, we should 
 consider the auxiliary field vertex operator $V^{(0)}_{H_{A X}}$ to 
 obtain gauge covariant result.

We do not need to calculate the interaction $[A_{\mu}, A_{\nu}]^2$ 
which can be generated after the integration of the auxiliary field $H_{\mu \nu}$
 in the Lagrangian, at the string level. Such interactions must be carefully 
extracted from those presented in the previous subsection.

We summarize all the possible vertex insertions that survives in the 
zero-slope limit in appendix A.
Any other amplitudes containing one graviphoton and open string vertex 
operators vanish in the zero-slope limit or are not consistent with the 
auxiliary field Lagrangian.

\subsection{Graviphoton effect}
Before calculating the corrections explicitly, we will examine the 
tensor structure of ${\cal F}^{\alpha\beta ij}$.
Since the space-time spinor indices $\alpha$ and the $R$-symmetry indices
$i$ are independent, we can classify the deformations as follows:
${\cal F}^{[\alpha\beta][ij]}$, ${\cal F}^{(\alpha\beta)[ij]}$,
${\cal F}^{[\alpha\beta](ij)}$ and ${\cal F}^{(\alpha\beta)(ij)}$.
We call these as (S,S), (S,A), (A,S), (A,A) type respectively. 
The general background contains all of these types simultaneously. 
It may be better to investigate each type
of deformation separately. In the following, we consider only the (S,S) type of the 
graviphoton background $\mathcal{F}^{\alpha \beta ij} = \mathcal{F}^{(\alpha \beta) 
(ij)}$. As we will see, this type of background corresponds to 
the graviphoton field strength 
that induces non(anti)commutative $\mathcal{N} = 1$ superspace 
$\{\theta^{\alpha}, \theta^{\beta}\} = C^{\alpha \beta}$ and the 
non-singletly deformed $\mathcal{N} = 2$ harmonic superspace $\{\theta^{i 
\alpha}, \theta^{j \beta}\} = C^{\alpha \beta} b^{ij}$. 
The vertex operator for the graviphoton field strength contains two 
internal spin fields $S_{i}, S_{j}$. These internal spin fields, when inserted on 
the disk without any other internal spin fields, generates anti-symmetric tensor 
$\varepsilon_{ij}$ through the correlator $\langle S_i (z) S_j (w) 
\rangle \sim \varepsilon_{ij}$.
When this anti-symmetric tensor is contracted with the graviphoton field 
strength $\mathcal{F}^{(\alpha \beta) (ij)}$, it gives vanishing amplitude. 
So, to obtain the non-vanishing amplitudes, at least one fermion vertex 
operator should 
be inserted in addition to the one graviphoton vertex operator. The 
cancellation condition of the $\phi_3$-charge implies that the smallest number of fermion insertion is actually 
two. Let us examine all the possible amplitudes including two fermion 
vertex operators below.
\\
\\
\textbullet \ 
\underline{$ \langle \! \langle V_{\Lambda} V_{\overline{\Lambda}} 
V_{\bar{\varphi}} V_{\mathcal{F}} \rangle \! \rangle + \langle \!
\langle 
V_{\Lambda} 
V_{\overline{\Lambda}} V_{H_{A \bar{\varphi}}} V_{\mathcal{F}} \rangle 
\! \rangle $}\\
\\
The first example of the amplitudes is $\langle \! \langle V_{\Lambda} 
V_{\overline{\Lambda}} V_{\bar{\varphi}} V_{\mathcal{F}} 
\rangle \! \rangle$. We should assign the picture 
number to each vertex operators adequately and evaluate the correlators. 
The amplitude is
\begin{eqnarray}
& & \langle \! \langle V^{(-1/2)}_{\Lambda} (p_1) 
V^{(-1/2)}_{\overline{\Lambda}} (p_2) V^{(0)}_{\bar{\varphi}} (p_3) 
V^{(-1/2,-1/2)}_{\mathcal{F}} \rangle \! \rangle_{\mathrm{(S,S)}} \nonumber \\
& & = \frac{1}{2 \pi^2 \alpha'^2} \frac{1}{k g^2_{\mathrm{YM}}} (2 \pi \alpha')
^{3} (2i)  \mathrm{tr} \left[ \Lambda^{\gamma k} (p_1) \overline{\Lambda}
_{\dot{\delta} l} (p_2) \bar{\varphi} (p_3) \right] \mathcal{F}^{(\alpha 
\beta) (ij)} \nonumber \\
& & \qquad \times \int \! \frac{\prod_j d y_j}{d V_{\mathrm{CKG}}} 
\langle e^{-\frac{1}{2} \phi (y_1) } e^{-\frac{1}{2} \phi (y_2) } 
e^{-\frac{1}{2} \phi (z) } e^{-\frac{1}{2} \phi (\bar{z}) } \rangle 
 \langle S_k (y_1) S^l (y_2) S_i (z) S_j (\bar{z}) \rangle \nonumber \\
& & \qquad \times i (2 \pi \alpha')^{\frac{1}{2}} p_{3 \mu} \langle S_{\gamma} (y_1)
S^{\dot{\delta}} (y_2) \psi^{\mu} (y_3) S_{\alpha} (z) S_{\beta} 
(\bar{z}) \rangle \nonumber \\
& & \qquad \times \langle S^{(-)} (y_1) S^{(+)} (y_2) \overline{\Psi} (y_3) 
 S^{(-)} (z) S^{(-)} (\bar{z}) \rangle  \left\langle \prod_{j=1}^3  e^{i 
					\sqrt{2 \pi \alpha'} p_j \cdot X 
					(y_j)} \right\rangle.
\end{eqnarray}
Here the symbol ``(S,S)'' means that we extract only the non-zero part 
from the correlator when it is contracted with the (S,S) type of the graviphoton
field strength $\mathcal{F}^{(\alpha \beta) (ij)}$.
After using the effective rules and also the massless condition, we see
\begin{eqnarray}
& & \langle \! \langle V^{(-1/2)}_{\Lambda} (p_1) 
V^{(-1/2)}_{\overline{\Lambda}} (p_2) V^{(0)}_{\bar{\varphi}} (p_3) 
V^{(-1)}_{\mathcal{F}} \rangle \! \rangle_{\mathrm{(S,S)}} \nonumber \\
& & = - \frac{1}{2 \pi^2 \alpha'^2}
\frac{1}{k g^2_{\mathrm{YM}}} (2 \pi \alpha')^{\frac{7}{2}} (2 i^2) \frac{1}{\sqrt{2}}
(\sigma^{\mu})_{\alpha}^{\ \dot{\delta}} \varepsilon_{\gamma \beta} 
e^{-\frac{1}{4} \pi i} 
\cdot I \cdot  \mathrm{tr} \left[ \Lambda_{\beta j}
(p_1) \overline{\Lambda}_{\dot{\delta} j} (p_2) p_{3 \mu} \bar{\varphi} (p_3)
 \right] \mathcal{F}^{(\alpha \beta) (ij)}. \nonumber \\
\end{eqnarray}
Here, the overall phase which comes from the cocycle factor and spin 
fields \cite{KoLeLeSaWa} is explicitly written.
The $SL(2,\mathbf{R})$ invariance is used to fix the positions
to $y_1 \to \infty, z \to i, \bar{z} \to - i$ \cite{Billo-1/2}. 
$I$ is the world sheet integral and is evaluated as
\begin{eqnarray}
I = 
%\int^{y_1}_{- \infty} \! d y_2 \ \int^{y_2}_{- \infty} \! d y_3 \ 
%\frac{(z - \bar{z})^2}{(y_2 - z) (y_2 - \bar{z} (y_3 - z) (y_3 - 
%\bar{z}))} \longrightarrow 
\int^{\infty}_{- \infty} \! d y_2 \ \int^{y_2}_{- \infty} \! d y_3 \ 
\frac{(2i)^2}{(y_2^2 + 1) (y_3^2 + 1)} = (2i)^2 \frac{\pi^2}{2}.
\label{WS-integral}
\end{eqnarray}
After all, the resulting amplitude is
\begin{eqnarray}
& &  \langle \! \langle V^{(-1/2)}_{\Lambda} (p_1) 
V^{(-1/2)}_{\overline{\Lambda}} (p_2) V^{(0)}_{\bar{\varphi}} (p_3) 
V^{(-1/2,-1/2)}_{\mathcal{F}} \rangle \! \rangle_{\mathrm{(S,S)}} \nonumber \\
& & \qquad = - \frac{2}{\sqrt{2}}
\frac{1}{k g^2_{\mathrm{YM}}} \mathrm{tr}
\left[ \Lambda_{\alpha i} (p_1) 
 \overline{\Lambda}_{\dot{\alpha} j} (p_2) (\sigma^{\mu})_{\beta}^{\ 
 \dot{\alpha}} i p_{3 \mu} \bar{\varphi} (p_3) \right] C^{(\alpha \beta) 
(ij)}, 
\end{eqnarray}
where we have defined $C^{(\alpha \beta) (ij)} \equiv - 4 \pi^2 
e^{\frac{1}{4} \pi i} (2 \pi 
\alpha')^{\frac{3}{2}} \mathcal{F}^{(\alpha \beta) (ij)}$.
By adding another color ordered contribution, we find that the 
amplitude is reproduced by the following interaction:
\begin{eqnarray}
\mathcal{L} = - \frac{1}{\sqrt{2}} \frac{1}{k g^2_{\mathrm{YM}}} \mathrm{tr} 
\left[ C^{(\alpha \beta) (ij)} \left\{ \partial_{\mu} \bar{\varphi} (x),
(\sigma^{\mu})_{\alpha \dot{\alpha}} 
\overline{\Lambda}^{\dot{\alpha}}_{\ i} (x) \right\} \Lambda_{\beta j} 
(x) \right].
\end{eqnarray}
This result contains derivative of the adjoint scalar which 
originates from the zero-ghost picture vertex operator $V^{(0)}_{\bar{\varphi}}$. 
As we noticed before, the auxiliary field amplitude $\langle \! \langle V_{\Lambda} 
V_{\overline{\Lambda}} V_{H_{A \bar{\varphi}}} V_{\mathcal{F}} 
\rangle \! \rangle$
 also contributes. In a similar way, the auxiliary 
 field contribution is evaluated to be
\begin{eqnarray}
& &\langle \! \langle V^{(-1/2)}_{\Lambda} (p_1) 
V^{(-1/2)}_{\overline{\Lambda}} (p_2) V^{(0)}_{H_{A \bar{\varphi}}} 
(p_3) V^{(-1/2,-1/2)}_{\mathcal{F}} \rangle \! \rangle_{(S,S)}
\nonumber \\
& & \qquad \qquad =  \frac{2}{\sqrt{2}} \frac{1}{k g^2_{\mathrm{YM}}} 
\mathrm{tr} \left[ \Lambda_{\alpha i} (p_1) \overline{\Lambda}_{\dot{\alpha}j}
(p_2) (\sigma^{\mu})_{\beta}^{\ \dot{\alpha}} 
{H_{A \bar{\varphi}}}_{\mu} (p_3) \right] C^{(\alpha \beta) (ij)}.
\end{eqnarray}
%\begin{eqnarray}
%\mathcal{L} = - \frac{ \sqrt{2}}{k g^2_{\mathrm{YM}}} 
%\mathrm{tr} \left[ C^{(\alpha \beta)(ij)} \left\{
%{H_{A \bar{\varphi}}}_{\mu} (x), (\sigma^{\mu})_{\alpha \dot{\alpha}} 
%\overline{\Lambda}^{\dot{\alpha}}_{\ i} (x) \right\} \Lambda_{\beta j} (x) \right].
%\nonumber \\
%\end{eqnarray}
After adding other inequivalent color order and multiplying the symmetric 
factor, we find the sum of the above two interactions is written as
\begin{eqnarray}
\mathcal{L} = - \frac{1}{\sqrt{2}} \frac{1} {k g^2_{\mathrm{YM}}} \mathrm{tr} 
\left[ C^{(\alpha \beta) (ij)} \left\{  \partial_{\mu} \bar{\varphi} (x)
+ {H_{A \bar{\varphi}}}_{\mu} (x), (\sigma^{\mu})_{\alpha \dot{\alpha}} 
\overline{\Lambda}^{\dot{\alpha}}_{\ i} (x) \right\} \Lambda_{\beta j} 
(x) \right].
\end{eqnarray}
\\
\textbullet \ \underline{$\langle \! \langle V_{A} V_{\overline{\Lambda}} V_{\overline{\Lambda}}
 V_{\mathcal{F}} \rangle \! \rangle + \langle \! \langle V_{H} V_{\overline{\Lambda}} 
 V_{\overline{\Lambda}} V_{\mathcal{F}} \rangle \! \rangle$}\\
\\
The next possible amplitude that can survive is $\langle \! \langle V_{V_A} 
V_{\overline{\Lambda}} V_{\overline{\Lambda}}  V_{\mathcal{F}} \rangle 
\! \rangle$, which is given by
\begin{eqnarray}
& & \langle \! \langle V^{(0)}_{A} (p_1) V^{(-1/2)}_{\overline{\Lambda}} (p_2)
V^{(-1/2)}_{\overline{\Lambda}} (p_3) V^{(-1/2,-1/2)}_{\mathcal{F}} \rangle \! 
\rangle_{\mathrm{(S,S)}} \nonumber \\
& & \qquad = \frac{1}{2 \pi^2 \alpha'^2} \frac{1}{k g^2_{\mathrm{YM}}} (2i)
(2 \pi \alpha')^{3} \mathrm{tr} \left[
A_{\mu} (p_1) \overline{\Lambda}_{\dot{\alpha} k} (p_2)
\overline{\Lambda}_{\dot{\beta} l} (p_3) \right] \mathcal{F}^{(\alpha 
\beta) (ij)} \int \! \frac{\prod_j d y_j}{d V_{\mathrm{CKG}}} \nonumber \\
& & \qquad \qquad \times \langle e^{- \frac{1}{2} \phi (y_2)} e^{- \frac{1}{2} \phi (y_3)}
 e^{- \frac{1}{2} \phi (z)} e^{- \frac{1}{2} \phi (\bar{z})} 
 \rangle  \langle S^k (y_2) S^l (y_3) S_i (z) S_j (\bar{z}) 
 \rangle \nonumber \\ 
& & \qquad \qquad \times \langle S^{(+)} (y_2) S^{(+)} (y_3) S^{(-)} (z) S^{(-)} (\bar{z}) 
 \rangle \nonumber \\
& & \qquad \qquad \times i (2 \pi \alpha')^{\frac{1}{2}} p_{1 \nu} \langle 
 \psi^{\nu} \psi^{\mu} (y_1) S^{\dot{\alpha}} (y_1) S^{\dot{\beta}} (y_2)
S_{\alpha} (z) S_{\beta} (\bar{z}) \rangle  \left\langle \prod_{j=1}^3 e^{i \sqrt{2 
\pi \alpha'} p_j \cdot X (y_j) } 
\right\rangle  \nonumber \\
& & = \frac{1}{2 \pi^2 \alpha'^2} \frac{1}{k g^2_{\mathrm{YM}}} (2 i^2)
(4 \pi^2 \alpha'^2) \cdot I \cdot \frac{1}{2} e^{- \frac{1}{4} \pi i} \mathrm{tr} \left[
(\sigma^{\mu \nu})_{\alpha \beta} p_{1 \mu} A_{\nu} (p_1) \overline{\Lambda}_{\dot{\alpha}i}
(p_2) \overline{\Lambda}_{\dot{\beta} i} (p_3) \right] (2 \pi \alpha')^{\frac{3}{2}} \mathcal{F}^{(\alpha \beta) 
(ij)} . \label{ALLF} \nonumber \\
\end{eqnarray}
Here we have introduced the Lorentz generators $\sigma_{\mu \nu} = \frac{1}{4} \left(
\sigma_{\mu} \bar{\sigma}_{\nu} - \sigma^{\nu} \bar{\sigma}_{\mu} \right)$.
%The world sheet integral $I'$ is given by $\frac{1}{2i} I$, where $I$ is 
% evaluated in (\ref{WS-integral}).
The world sheet integral $I$ is given in (\ref{WS-integral}).
After all, this amplitude is evaluated as 
\begin{eqnarray}
& & \langle \! \langle V^{(0)}_{A} (p_1) V^{(-1/2)}_{\overline{\Lambda}} (p_2)
V^{(-1/2)}_{\overline{\Lambda}} (p_3) V^{(-1/2,-1/2)}_{\mathcal{F}} \rangle \! 
\rangle_{\mathrm{(S,S)}} \nonumber \\
& & =  \frac{4 \pi^2 i}{k g^2_{\mathrm{YM}}}  \mathrm{tr} \left[
(\sigma^{\mu \nu})_{\alpha \beta} p_{1 \mu} A_{\nu} (p_1) 
\overline{\Lambda}_{\dot{\alpha} i} (p_2) \overline{\Lambda}^{\dot{\alpha}}_{\ 
 j} (p_3) \right] (2\pi \alpha')^{\frac{3}{2}} e^{- \frac{1}{4} \pi i} \mathcal{F}^{(\alpha \beta) (ij)}.
\end{eqnarray}
The auxiliary field insertion $\langle \! \langle V_{H} V_{\overline{\Lambda}} 
V_{\overline{\Lambda}} V_{\mathcal{F}} \rangle \! \rangle$ also contributes.
The calculation of this amplitude 
%for the $\langle \! \langle V_{H} V_{\overline{\Lambda}} 
%V_{\overline{\Lambda}} V_{\mathcal{F}} \rangle \! \rangle$ part
is essentially the same as (\ref{ALLF}). The result is
\begin{eqnarray}
& & \langle \! \langle V^{(0)}_{H} (p_1) V^{(-1/2)}_{\overline{\Lambda}} (p_2)
V^{(-1/2)}_{\overline{\Lambda}} (p_3) V^{(-1/2,-1/2)}_{\mathcal{F}} \rangle \! 
\rangle_{\mathrm{(S,S)}} \nonumber \\
& & \qquad \qquad = \frac{1}{4} \frac{1}{k g^2_{\mathrm{YM}}} \mathrm{tr} 
\left[ (\sigma^{\mu \nu})_{\alpha \beta} H_{\mu \nu} (p_1) \overline{\Lambda}_{\dot{
\alpha} i} (p_1) \overline{\Lambda}^{\dot{\alpha}}_{j} (p_3) 
\right] C^{(\alpha \beta) (ij)}. 
\end{eqnarray}
By adding another color ordered amplitude and making 
the phase shift of $\overline{\Lambda}$, we obtain the 
graviphoton induced Lagrangian
\begin{eqnarray}
\mathcal{L} &=& 
%\frac{i}{k g^2_{\mathrm{YM}}} \mathrm{tr} \left[(\sigma^{\mu \nu})_{\alpha \beta} \partial_{\mu}
%A_{\nu} (x) \overline{\Lambda}_{\dot{\alpha} i} (x) 
%\overline{\Lambda}_{\dot{\beta} j} (x) \varepsilon^{\dot{\alpha} 
%\dot{\beta}} C^{(\alpha \beta) (ij)}
%\right] \nonumber \\
%& & \qquad \qquad - \frac{1}{4} \frac{1}{k g^2_{\mathrm{YM}}} \mathrm{tr} \left[(\sigma^{\mu \nu})_{\alpha \beta} H_{\mu \nu} (x) \overline{\Lambda}_{\dot{\alpha} i} (x) 
%\overline{\Lambda}_{\dot{\beta} j} (x) \varepsilon^{\dot{\alpha} \dot{\beta}} C^{(\alpha \beta) (ij)}
%\right] \nonumber \\
%& & 
= - \frac{i}{2} \frac{1}{k g^2_{\mathrm{YM}}} \mathrm{tr}
\left[ \left\{ (\partial_{\mu} A_{\nu} (x) - \partial_{\nu} A_{\mu} (x) ) - 
 \frac{i}{2} H_{\mu \nu} (x) \right\} \overline{\Lambda}_{\dot{\alpha} i} (x) 
\overline{\Lambda}^{\dot{\alpha}}_{\ j} (x) \right] C^{\mu \nu (ij)}.
\end{eqnarray}
Here we have defined $C^{\mu \nu (ij)} \equiv (\sigma^{\mu \nu})_{\alpha 
\beta} C^{(\alpha \beta) (ij)}$.
\\\\
\textbullet \ \underline{$\langle \! \langle V_{\bar{\varphi}} 
V_{\overline{\Lambda}} V_{\overline{\Lambda}} 
V_{\bar{\varphi}} V_{\mathcal{F}} \rangle \! \rangle$}\\
\\
The amplitude of the form $\langle \! \langle V_{\bar{\varphi}} 
V_{\overline{\Lambda}} V_{\overline{\Lambda}} 
V_{\bar{\varphi}} V_{\mathcal{F}} \rangle \! \rangle$ is also the
candidate for the non-vanishing amplitude:
\begin{eqnarray}
& & \langle \! \langle V^{(0)}_{\bar{\varphi}} (p_1) V^{(-1/2)}_{\Lambda} (p_2)
V^{(-1/2)}_{\Lambda} (p_3) V^{(0)}_{\bar{\varphi}} (p_4) 
V^{(-1/2,-1/2)}_{\mathcal{F}} \rangle \! \rangle \nonumber \\
& & = \frac{1}{2 \pi^2 \alpha'^2} \frac{1}{k g^2_{\mathrm{YM}}} (2 \pi 
 \alpha')^{\frac{7}{2}} (2 i)^2 \mathrm{tr} \left[ \bar{\varphi} (p_1) 
\Lambda^{k \gamma} (p_2) \Lambda^{l \delta} (p_3) \bar{\varphi} (p_4) \right]
\mathcal{F}^{\alpha \beta ij} \nonumber \\
& & \times \int \!\! \frac{\prod_j dy_j}{d V_{\mathrm{CKG}}}
\langle e^{- \frac{1}{2} \phi (y_2)} e^{- \frac{1}{2} \phi (y_3)}
e^{- \frac{1}{2} \phi (z)} e^{- \frac{1}{2} \phi (\bar{z})} \rangle 
\langle S_k (y_2) S_l (y_3) S_i (z) S_j (\bar{z}) \rangle \nonumber \\
& & \qquad \times \left\langle \left( \partial \overline{Z} (y_1) 
+ i (2 \pi \alpha')^{\frac{1}{2}} p_1 \cdot \psi \overline{\Psi} (y_1)  \right)
S^{(-)} (y_2) S^{(-)} (y_3) S_{\gamma} (y_2) S_{\delta} (y_3) \right. 
\nonumber \\
& & \left. \qquad \times \left( \partial \overline{Z} (y_4) 
+ i (2 \pi \alpha')^{\frac{1}{2}} p_4 \cdot \psi \overline{\Psi} (y_4)  \right)
S^{(-)} (z) S^{(-)} (\bar{z}) S_{\alpha} (z) S_{\beta} (\bar{z}) \right\rangle
\left\langle \prod_{j=1}^4 e^{i \sqrt{2 \pi \alpha'} p_j \cdot X (y_j)} 
     \right\rangle. \nonumber \\
\end{eqnarray}
The $\partial \overline{Z} \partial \overline{Z}$ part and the cross 
terms does not contribute to the amplitude because the $\phi_3$-charge can 
not be canceled. The non-zero contribution comes from the  $p_1 \cdot \psi \overline{\Psi} 
\ p_2 \cdot \psi \overline{\Psi} $ part only. The correlator is reduced 
to the form
\begin{eqnarray}
& & i^2 (2 \pi \alpha') p_{1 \mu} p_{2 \nu} 
\langle \psi^{\mu} (y_1) S_{\gamma} (y_2) S_{\delta} (y_3) \psi^{\nu} (y_4)
S_{\alpha} (z) S_{\beta} (\bar{z}) \rangle \nonumber \\
& & \qquad \times \langle \overline{\Psi} (y_1) S^{(-)} (y_2) S^{(-)} (y_3)
\overline{\Psi} (y_4) S^{(-)} (z) S^{(-)} (\bar{z}) \rangle \nonumber \\
& & \qquad \times \langle e^{- \frac{1}{2} \phi (y_2)} e^{- \frac{1}{2} \phi (y_3)}
e^{- \frac{1}{2} \phi (z)} e^{- \frac{1}{2} \phi (\bar{z})} \rangle.
\end{eqnarray}
However this is higher $\alpha'$ order contribution so that it does not 
survive in the zero-slope limit in our scaling.
\\
\\
\textbullet \ \underline{
$\langle \! \langle V_{\overline{\Lambda}} V_{\overline{\Lambda}} V_{\mathcal{F}}
\rangle \! \rangle$}\\
\\
The amplitude is 
\begin{eqnarray}
& & \langle \! \langle V_{\overline{\Lambda}}^{(-1/2)} (p_1) V_{\overline{\Lambda}}^{(-1/2)}
(p_2) V_{\mathcal{F}}^{(-1/2,-1/2)} \rangle \! \rangle_{(S,S)} \nonumber \\
& & = \frac{1}{2 \pi^2 \alpha'^2} \frac{1}{k g^2_{\mathrm{YM}}} (2 \pi \alpha')^{\frac{5}{2}}
 \mathrm{tr} \left[  
\overline{\Lambda}_{\dot{\alpha} k} (p_1)
 \overline{\Lambda}_{\dot{\beta} l} (p_2) \right]  \mathcal{F}^{\alpha 
\beta ij} \nonumber \\
& & \times  \int \!\! \frac{\prod_j d y_j}{d V_{\mathrm{CKG}}} \langle 
 e^{- \frac{1}{2} \phi (y_1)} e^{- \frac{1}{2} \phi (y_2)} e^{ - 
 \frac{1}{2}} \phi (z) e^{- \frac{1}{2} \phi (\bar{z})} \rangle 
  \langle S^{\dot{\alpha}} (y_1) S^{\dot{\beta}} (y_2) 
S_{\alpha} (z) S_{\beta} (\bar{z}) \rangle \nonumber \\
& & \times \langle S^{(+)} (y_1) 
S^{(+)} (y_2) S^{(-)} (z) S^{(-)} (\bar{z}) \rangle 
 \langle S^k (y_1) S^l (y_2) S_i (z) S_j (\bar{z}) \rangle 
\left\langle \prod_{j=1}^3 e^{i \sqrt{2 \pi \alpha'} p_j \cdot X (y_j)} \right\rangle.
\nonumber \\
\end{eqnarray}
The effective rule for the four-dimensional spin field correlator is
\begin{eqnarray}
\langle S^{\dot{\alpha}} (y_1) S^{\dot{\beta}} (y_2) 
S_{\alpha} (z) S_{\beta} (\bar{z}) \rangle 
= \varepsilon^{\dot{\alpha} \dot{\beta}} \varepsilon_{\alpha \beta}
(y_1 - y_2)^{-\frac{1}{2}} (z - \bar{z})^{-\frac{1}{2}},
\end{eqnarray}
which gives vanishing contribution when contracted with the (S,S) type of graviphoton.
\\
\\
\textbullet \ \underline{$\langle \! \langle V_{H_{\varphi \bar{\varphi}}}
 V_{\overline{\Lambda}} V_{\overline{\Lambda}} V_{\mathcal{F}} \rangle \! 
 \rangle $}\\
\\
The amplitude is
\begin{eqnarray}
& & \langle \! \langle V^{(0)}_{H_{\varphi \bar{\varphi}}} (p_1) V^{(-1/2)}_{\overline{\Lambda}}
(p_2) V^{(-1/2)}_{\overline{\Lambda}} (p_3) V^{(-1/2,-1/2)}_{\mathcal{F}} 
\rangle \! \rangle \nonumber \\
& & = \frac{1}{2 \pi^2 \alpha'^2} \frac{1}{k g^2_{\mathrm{YM}}} (2 \pi 
 \alpha')^{\frac{7}{2}} (-i \sqrt{2}) \mathrm{tr} \left[ H_{\varphi \bar{\varphi}} (p_1) 
  \overline{\Lambda}_{\dot{\gamma}k} (p_2) 
  \overline{\Lambda}_{\dot{\delta}l} (p_3)  \right] \mathcal{F}^{\alpha 
\beta ij} \nonumber \\
& & \times \int \!\! \frac{\prod_j d y_j}{d V_{\mathrm{CKG}}} 
\langle e^{-\frac{1}{2} \phi (y_2)} e^{-\frac{1}{2} \phi (y_3)} e^{-\frac{1}{2} \phi (z)}
e^{-\frac{1}{2} \phi (\bar{z})} \rangle \nonumber \\
& & \times \langle S^k (y_2) S^l (y_3) S_i (z) S_j (\bar{z}) \rangle
\langle S^{\dot{\gamma} } (y_2) S^{\dot{\delta}} (\bar{z}) \rangle 
\langle S_{\alpha} (z) S_{\beta} (\bar{z}) \rangle \nonumber \\
& & \times \langle \Psi \overline{\Psi} (y_1) \ S^{(+)} (y_2) S^{(+)} (y_3)
S^{(-)} (z) S^{(-)} (\bar{z}) \rangle \left\langle \prod_{j=1}^3 e^{i 
					\sqrt{2 \pi \alpha'} p_j \cdot X 
					(y_j)} \right\rangle.
\end{eqnarray}
In this case, there is a factor $\varepsilon_{\alpha \beta}$ coming from 
the spin field correlator $\langle S_{\alpha} (z) S_{\beta} (\bar{z}) \rangle$. 
Thus when it is contracted with the (S,S) type of the graviphoton, this part gives vanishing result.
\\
\\
Altogether, the interaction term $\mathcal{L}_{(S,S)}$ in the Lagrangian 
induced by the (S,S) type of the graviphoton field strength at the 
lowest order is
\begin{eqnarray}
& & \mathcal{L}_{(S,S)} = - \frac{1}{\sqrt{2}} \frac{1}{k g^2_{\mathrm{YM}}} \mathrm{tr} 
 \left[ C^{(\alpha \beta) (ij)} \left\{  \partial_{\mu} \bar{\varphi} (x) 
+ {H_{A \bar{\varphi}}}_{\mu} (x) , (\sigma^{\mu})_{\alpha \dot{\alpha}} 
\overline{\Lambda}^{\dot{\alpha}}_{\ i} (x)  \right\} \Lambda_{\beta j} (x) 
\right] \nonumber \\
& & \qquad \qquad \qquad -  \frac{i}{2} \frac{1 }{k g^2_{\mathrm{YM}}} \mathrm{tr}
\left[ \left\{ (\partial_{\mu} A_{\nu} (x) - \partial_{\nu} A_{\mu} (x) ) - 
 \frac{i}{2} H_{\mu \nu} (x) \right\} \overline{\Lambda}_{\dot{\alpha} i} (x) 
\overline{\Lambda}^{\dot{\alpha}}_{\ j} (x) \right] C^{\mu \nu (ij)}.
\nonumber \\
\end{eqnarray}

After integrating out the auxiliary fields, we find effective 
interaction terms are written as 
\begin{eqnarray}
\mathcal{L}_{(S,S)} &=& - \frac{1}{\sqrt{2}} \frac{1}{g^2_{\mathrm{YM}}} 
 \frac{1}{k} \mathrm{tr} \left[ C^{(\alpha \beta) (ij)}
\{ D_{\mu} \bar{\varphi}, (\sigma^{\mu})_{\alpha \dot{\alpha}} 
\overline{\Lambda}^{\dot{\alpha}}_{\ i}  \} \Lambda_{\beta j}
 \right] \nonumber \\
 & &  - \frac{i}{2} \frac{1}{g^2_{\mathrm{YM}}} \frac{1}{k} \mathrm{tr} 
\left[ F_{\mu \nu} \overline{\Lambda}_i 
\overline{\Lambda}_j C^{\mu \nu (ij)}  \right]  + \frac{1}{8} \frac{1}{g^2_{\mathrm{YM}}}
\frac{1}{k} \mathrm{tr} \left[ \overline{\Lambda}_i 
\overline{\Lambda}_j C^{\mu \nu (ij)} \overline{\Lambda}_k 
\overline{\Lambda}_l C_{\mu \nu}^{\ \ (kl)} \right]. \label{SS}
\end{eqnarray} 
If we consider the case that only the part $C^{\alpha \beta} \equiv C^{\alpha \beta 11} $ is
non-zero then we find the deformed Lagrangian
\begin{eqnarray}
\mathcal{L}_c = \mathcal{L}^{\mathcal{N} = 2}_{\mathrm{SYM}} + \mathcal{L}_{(S,S)} 
\label{(S,S)}
\end{eqnarray}
precisely coincides with the one constructed in 
the $\mathcal{N} = 1/2$ superspace with the Moyal product \cite{ArItOh1}\footnote{
Here, compared with the Lagrangian in \cite{ArItOh1}, we have rescaled 
$A_{\mu} \rightarrow \frac{1}{g^2_{\mathrm{YM}}} A_{\mu}$, 
$(A,\bar{A}) \rightarrow \frac{1}{g^2_{\mathrm{YM}}} (A, \bar{A})$, 
$C^{\alpha \beta} \rightarrow \frac{1}{g^2_{\mathrm{YM}}} C^{\alpha 
\beta} $.}
\begin{eqnarray}
\mathcal{L} &=& \frac{1}{g^2_{\mathrm{YM}}} \frac{1}{k} \mathrm{tr}
\left[
- \frac{1}{4} F_{\mu \nu} F^{\mu \nu} - \frac{1}{4} F_{\mu \nu}
\tilde{F}^{\mu \nu} - i \bar{\lambda}_{\dot{\alpha}} 
 (\bar{\sigma}^{\mu})^{\dot{\alpha} \alpha} D_{\mu} \lambda_{\alpha} 
- i \bar{\psi}_{\dot{\alpha}} (\bar{\sigma}^{\mu})^{\dot{\alpha} \alpha} 
D_{\mu} \psi_{\alpha} \right. \nonumber \\
& & \left. \qquad \qquad \qquad - (D_{\mu} \bar{A}) (D^{\mu} A) - i \sqrt{2} [\bar{A}, \psi^{\mu}] \lambda_{\alpha} 
- i \sqrt{2} [A, \bar{\psi}_{\dot{\alpha}}] \bar{\lambda}^{\dot{\alpha}} 
 - \frac{1}{2} [A, \bar{A}]^2 
\right] \nonumber \\
& & + \frac{1}{g^2_{\mathrm{YM}}} \frac{1}{k} \mathrm{tr}
\left[
- \frac{i}{2} C^{\mu \nu} F_{\mu \nu} \bar{\lambda}_{\dot{\alpha}} 
\bar{\lambda}^{\dot{\alpha}} + \frac{1}{8} |C|^2 (\bar{\lambda}_{\dot{\alpha}} \bar{\lambda}^{\dot{\alpha}})^2
 - \frac{\sqrt{2}}{2} C^{\alpha \beta} \{ D_{\mu} \bar{A}, 
(\sigma^{\mu})_{\alpha \dot{\alpha}} \bar{\lambda}^{\dot{\alpha}} \} 
\psi_{\beta} \right]. \nonumber \\ 
\end{eqnarray}
Here, we have defined $\bar{\varphi} = \bar{A}$, $\Lambda^1 \equiv \lambda, \overline{\Lambda}_1 
\equiv \bar{\lambda}, \Lambda_1 \equiv \psi, \overline{\Lambda}^1 = 
\bar{\psi}$.
%\begin{eqnarray}
%F_{\mu \nu} = \partial_{\mu} A_{\nu} - \partial_{\nu} A_{\mu} + i 
% [A_{\mu}, A_{\nu}], \qquad D_{\mu} \lambda_{\alpha} = \partial_{\mu} 
%\lambda_{\alpha} + i [A_{\mu} , \lambda_{\alpha}].
%\end{eqnarray}
Actually, this (S,S) type of the R-R background 
$\mathcal{F}^{(\alpha \beta) (11)}$ corresponds to 
the graviphoton vertex operator which induces the non-anticommutativity 
in the $\mathcal{N} = 1$ superspace \cite{Billo-1/2},
\begin{eqnarray}
V_{\mathcal{F}}^{(-1/2,-1/2)} (z, \bar{z}) \!\! 
 = \!\! (2 \pi \alpha')  \mathcal{F}^{\alpha \beta (11)} 
 \left[ S_{\alpha} (z) S^{(---)} e^{- \frac{1}{2} \phi (z)} 
  S_{\beta} (\bar{z}) S^{(---)} (\bar{z}) e^{ - \frac{1}{2} 
  \phi (\bar{z})} \right] \label{N=1graviphoton}.
\end{eqnarray}
It is worth to mention that deformation to the super Yang-Mills 
action in the background (\ref{N=1graviphoton}) terminates at the 
quadratic order in $\mathcal{F}$ though it is not the case for the 
general graviphoton background. The $\mathcal{F}^2$ term appeared in the 
(\ref{(S,S)}) is the only possible one.

We thus conclude that the $\mathcal{N} = 2$ super Yang-Mills theory defined 
on the $\mathcal{N} = 1/2$ superspace 
$\{\theta^{\alpha}, \theta^{\beta} \} = C^{\alpha \beta}$ is the 
effective theory of the D3-branes in the background graviphoton field 
of the type (\ref{N=1graviphoton}). 
The effective Lagrangian preserves only a part 
of the original supersymmetry but the canonical gauge invariance is intact
\cite{ArItOh1}. Note that the background corresponding to the Lagrangian 
(\ref{(S,S)}) is self-dual and does not receive the gravitational 
back-reaction.

The general type of the graviphoton background $\mathcal{F}^{(\alpha 
\beta) (ij)}$ seems to correspond to the non-singlet deformation 
$\{\theta^{i \alpha }, 
\theta^{j \beta } \} = C^{\alpha \beta} b^{ij} $ of 
$\mathcal{N} = 2$ harmonic superspace. In fact, 
decomposing the scaled graviphoton field strength
$C^{(\alpha \beta) (ij)}$ into 
the the form 
$C^{\alpha \beta} b^{ij}$ and setting $b^{ij}=\delta^i_1 \delta^j_1$
\cite{ArItOh5}. In fact, for $b^{ij}$ satisfying
$b^{ij} b_{ij} = 0$, the exact deformed 
$\mathcal{N} = 2$ abelian gauge theory was obtained in \cite{Castro}, 
which is of the form
\begin{eqnarray}
\mathcal{L} &=& - \partial_{\mu} \varphi \partial^{\mu} \bar{\varphi}
- \frac{1}{4} F_{\mu \nu} (F^{\mu \nu} + \tilde{F}^{\mu \nu}) 
- i \Lambda^{\alpha i} (\sigma^{\mu})_{\alpha \dot{\alpha}} \partial_{\mu}
\overline{\Lambda}^{\dot{\alpha}}_{\ i} \nonumber \\
& & + 4 \sqrt{2} i C^{\alpha \beta} b^{ij} (\sigma^{\mu})_{\alpha 
 \dot{\alpha}} \partial_{\mu} \bar{\varphi} \cdot \overline{\Lambda}^{\dot{\alpha}}
_{\ i} \Lambda_{\beta j} - 2 C^{\mu \nu} b^{ij} F_{\mu \nu} \overline{\Lambda}_{\dot{\alpha}i}
\overline{\Lambda}^{\dot{\alpha}}_{\ j} - 2 C^{\mu \nu} 
b^{ij} C_{\mu \nu} b^{kl} (\overline{\Lambda}_i \overline{\Lambda}_j)
 (\overline{\Lambda}_k \overline{\Lambda}_l). \nonumber \\ \label{NS}
\end{eqnarray}
If we identify $C^{\alpha \beta} b^{ij} \equiv \frac{i}{4} C^{(\alpha 
\beta) (ij)}$, the non-singlet deformed Lagrangian (\ref{NS}) exactly 
agrees with the (S,S) type deformed theory (\ref{(S,S)}).
For the non-singlet case with $b^{ij} b_{ij} \not= 0$, we can show that the deformed Lagrangian (\ref{(S,S)}) 
agrees with that of \cite{Castro} at the first order in $bc$.
%It was shown that the induced interactions from the
%non-anticommutativity of 
%harmonic 
%superspace $\{\theta^{\alpha i}, \theta^{\beta j} \} = C^{\alpha 
%\beta} b^{ij}$ are of the structure $ C^{\alpha \beta} b_{ij}
%\Lambda_{\beta}^{\ j}  (\sigma^{\mu})_{\alpha \dot{\alpha}} \partial_{\mu}
%\bar{\varphi} \overline{\Lambda}^{\dot{\alpha} i}$, $C^{\alpha \beta} 
%F_{\alpha \beta} \overline{\Lambda}_i \overline{\Lambda}_j b^{ij}$, $C^{\alpha \beta} 
%C_{\alpha \beta} (b_{ij} \overline{\Lambda}^i \overline{\Lambda}^j)^2$. 
%We can see these interactions precisely correspond to our result 
%presented in (\ref{SS}).

\section{Conclusions and Discussion}

In this paper, we have written down the low-energy effective Lagrangian
of $\mathcal{N}=2$ supersymmetric gauge theory 
from the open superstring amplitudes in the graviphoton background.
The structure of the deformed action depends on the 
scaling condition of the background in the zero-slope limit. 
We have chosen that the deformation parameter
have the same dimension of the non-anticommutativity parameter of the 
superspace, {\it i.e.} the graviphoton polarization scales as $(2 \pi \alpha')
^{\frac{3}{2}} \mathcal{F} = \mathrm{fixed}$ in the zero-slope limit.

Compared with the deformation of $\mathcal{N} = 1$ 
super Yang-Mills theory
 \cite{Billo-1/2},
 where only finite number of graviphoton vertex 
operator insertion in the disk amplitude is allowed, 
arbitrary number of graviphoton 
vertex operators 
can be inserted in the disk amplitudes 
in the case of the deformation of $\mathcal{N} = 2$ theory.

We have discussed that 
the graviphoton field strength $\mathcal{F}^{\alpha \beta ij}$ can be 
classified 
into four types: (S,S), (S,A), (A,S) and (A,A) types. 
In the present work, we have investigated the (S,S) type deformation
in detail.
For $(S,S)$ type deformation,
we have shown
 that the $\mathcal{N} = 2$ super Yang-Mills theory 
defined on $\mathcal{N} 
= 1$ non(anti)commutative superspace is precisely equivalent to the 
effective theory of the D3-branes in the presence of the self-dual
$\mathcal{F}^{(\alpha \beta) (11)}$  graviphoton background. We also find 
that the deformed $\mathcal{N} = 2$ abelian gauge theory defined on 
non(anti)commutative harmonic superspace 
$\{\theta^{i \alpha }, \theta^{j \beta }\} 
= C^{\alpha \beta} b^{ij}$ with $b^{ij} b_{ij} = 0$ \cite{Castro} is 
reproduced by the disk amplitude. 
On the other hand the deformations 
with parameters ${\cal F}^{[\alpha\beta](ij)}$ and 
${\cal F}^{(\alpha\beta)[ij]}$
do not correspond to the deformation of superspace.
These types of background would give new types of deformation of
$\mathcal{N}=2$ theory. In \cite{Billo-N=2}, the (S,A) type of 
the graviphoton $\mathcal{F}^{(\alpha \beta) [ij]}$ has been discussed.
They considered the scaling $(2 \pi \alpha')^
{\frac{1}{2}} \mathcal{F} = \mathrm{fix}$, which is different from ours. 
In this scaling, the non-zero contribution comes from only two amplitudes
\begin{eqnarray}
\langle \! \langle V_A V_{\bar{\varphi}} V_{\mathcal{F}} 
\rangle \! \rangle, \qquad 
\langle \! \langle V_{H} V_{\bar{\varphi}} V_{\mathcal{F}} 
\rangle \! \rangle.
\end{eqnarray}
Thus the effective Lagrangian after integrating out the auxiliary 
field is shown to be
\begin{eqnarray}
\mathcal{L}_c = \mathcal{L}_{\mathrm{SYM}}^{\mathcal{N} = 2} + \mathcal{L}',
\end{eqnarray}
where the induced interaction $\mathcal{L}'$ is simply given by
\begin{eqnarray}
\mathcal{L}' = \frac{1}{k g^2_{\mathrm{YM}}} \mathrm{tr} 
\left[F_{\mu \nu} \bar{\varphi} \tilde{C}^{\mu \nu} 
+ (\bar{\varphi} \tilde{C}^{\mu \nu})^2 \right].
\end{eqnarray}
We can also study the $(A,A)$ type deformation, which is expected to 
correspond to the singlet deformation of $\mathcal{N}=2$ harmonic
superspace \cite{FeIvLeSoZu}. 
But it is easily found that this theory  includes
divergence such as ${1\over (2\pi\alpha')^2}{\rm tr}\bar{\varphi} C$
in the zero-slope limit.
Therefore it is necessary to consider renormalization or back reactions 
in this type of deformation.
We will examine 
 deformed $\mathcal{N} = 2$ gauge theories corresponding to 
these types of graviphoton backgrounds
in a forthcoming paper.
We can also extend the present construction to 
deformed
${\cal N}=4$ supersymmetric gauge theory, 
where the deformed Lagrangian in the $\mathcal{N}=1/2$ superspace 
is known \cite{Im2}.
String theory calculation would provide more general type deformation
of $\mathcal{N}=4$ theory.
This subject will be also discussed in a separate paper.

\subsection*{Acknowledgments}
The authors would like to thank Y.~Kobayashi for useful discussions.
The work of K.~I. is  supported in part by the Grant-in-Aid for Scientific 
Research No. 18540255 from Ministry of Education, Science, 
Culture and Sports of Japan. S.~S. is supported by the Interactive Research 
Center of Science at Tokyo Institute of Technology.

\begin{appendix}
\section{Classification of possible disk amplitudes with one graviphoton
 insertion}
All possible contributions from the disk amplitudes including one
 graviphoton vertex operator are summarized as follows:
\renewcommand{\theenumi}{(\Roman{enumi})}
\begin{enumerate}
\item $ \displaystyle \langle \! \langle V_{\bar{\varphi}}^{(-1)} 
 V_{\mathcal{F}}^{(-1/2,-1/2)} \rangle \! \rangle$
\item $ \displaystyle \langle \! \langle V_{A}^{(0)} V_{\bar{\varphi}}^{(-1)} V_{\mathcal{F}}^{(-1/2,-1/2)} 
 \rangle \! \rangle,    \langle \! \langle V_{H}^{(0)} V_{\bar{\varphi}}^{(-1)} V_{\mathcal{F}}^{(-1/2,-1/2)} 
\rangle \! \rangle$
\item $ \displaystyle \langle \! \langle V_{A}^{(0)} V_{A}^{(0)} V_{\bar{\varphi}}^{(-1)}
 V_{\mathcal{F}}^{(-1/2,-1/2)} \rangle \! \rangle,   \langle \! \langle V_{A}^{(0)} V_{H}^{(0)}
 V_{\bar{\varphi}}^{(-1)} V_{\mathcal{F}}^{(-1/2,-1/2)} \rangle \! 
      \rangle, \\ \qquad \langle \! \langle V_{H}^{(0)} V_{A}^{(0)} V_{\bar{\varphi}}^{(-1)} 
V_{\mathcal{F}}^{(-1/2,-1/2)} \rangle \! \rangle \ 
, \langle \! \langle V_{H}^{(0)} V_{H}^{(0)} V_{\bar{\varphi}}^{(-1)} 
V_{\mathcal{F}}^{(-1/2,-1/2)} \rangle \! \rangle$ 
\item $ \displaystyle \langle \! \langle V^{(0)}_{\varphi} V^{(0)}_{\bar{\varphi}} 
 V^{(-1)}_{\bar{\varphi}} V^{(-1/2,-1/2)}_{\mathcal{F}} \rangle \! \rangle 
, \ \langle \! \langle V^{(0)}_{H_{A \varphi}} V^{(0)}_{\bar{\varphi}} 
V^{(-1)}_{\bar{\varphi}} V^{(-1/2,-1/2)}_{\mathcal{F}} \rangle \! \rangle, \\
\qquad  \langle \! \langle V^{(0)}_{\varphi} V^{(0)}_{H_{A \bar{\varphi}}} 
 V^{(-1)}_{\bar{\varphi}} V^{(-1/2,-1/2)}_{\mathcal{F}} \rangle \! \rangle 
,\ \langle \! \langle V^{(0)}_{H_{A \varphi}} V^{(0)}_{H_{A \bar{\varphi}}} 
 V^{(-1)}_{\bar{\varphi}} V^{(-1/2,-1/2)}_{\mathcal{F}} \rangle \! \rangle$
\item $ \displaystyle \langle \! \langle V^{(0)}_{H_{\varphi 
\bar{\varphi}}} V^{(0)}_{H_{\varphi \bar{\varphi}}} 
V^{(-1)}_{\bar{\varphi}} V^{(-1/2,-1/2)}_{\mathcal{F}} \rangle \! \rangle $
\item $ \displaystyle \langle \! \langle V^{(0)}_A V^{(0)}_{H_{\varphi \bar{\varphi}}} 
 V^{(-1)}_{\bar{\varphi}} V^{(-1/2,-1/2)}_{\mathcal{F}} \rangle \! \rangle 
, \ \langle \! \langle V^{(0)}_{H} V^{(0)}_{H_{\varphi \bar{\varphi}}} V^{(-1)}_{\bar{\varphi}}
V^{(-1/2,-1/2)}_{\mathcal{F}} \rangle \! \rangle$
\item $ \displaystyle \langle \! \langle V^{(-1/2)}_{\Lambda} 
V^{(-1/2)}_{\overline{\Lambda}}  V^{(0)}_{\bar{\varphi}} 
V^{(-1/2,-1/2)}_{\mathcal{F}} \rangle \! \rangle , \
\langle \! \langle V^{(-1/2)}_{\Lambda} V^{(-1/2)}_{\overline{\Lambda}} 
 V^{(0)}_{H_{A \bar{\varphi}}} V^{(-1/2,-1/2)}_{\mathcal{F}} \rangle \! 
 \rangle$
\item $ \displaystyle \langle \! \langle V^{(0)}_{\bar{\varphi}} V^{(-1/2)}_{\Lambda} 
 V^{(-1/2)}_{\Lambda} V^{(0)}_{\bar{\varphi}} V^{(-1/2,-1/2)}_{\mathcal{F}} 
 \rangle \! \rangle$
\item $ \displaystyle \langle \! \langle V_{\bar{\Lambda}}^{(-1/2)} V_{\bar{\Lambda}}^{(-1/2)} 
 V^{(-1/2,-1/2)}_{\mathcal{F}} \rangle \! \rangle$
\item $ \displaystyle \langle \! \langle V^{(0)}_A V^{(-1/2)}_{\overline{\Lambda}} V^{(-1/2)}_{\overline{\Lambda}}
V^{(-1/2,-1/2)}_{\mathcal{F}} \rangle \! \rangle ,\
\langle \! \langle V^{(0)}_{H} V^{(-1/2)}_{\overline{\Lambda}} V^{(-1/2)}_{\overline{\Lambda}}
V^{(-1/2,-1/2)}_{\mathcal{F}} \rangle \! \rangle$
\item $ \displaystyle \langle \! \langle V^{(0)}_{H_{\varphi \bar{\varphi}}} 
 V^{(-1/2)}_{\overline{\Lambda}} V^{(-1/2)}_{\overline{\Lambda}} 
V^{(-1/2,-1/2)}_{\mathcal{F}} \rangle \! \rangle$.
\end{enumerate}

\section{Effective rules}
The ten-dimensional correlator is computed by decomposing it into 
the four-dimensional part and the internal part \cite{Billo-1/2}. The 
effective rules that enable one to calculate each part separately, is 
derived by the general formula \cite{KoLeLeSaWa}. For the 
four-dimensional spin field, we have
\begin{eqnarray}
\langle S_{\alpha} (z) S_{\beta} (\bar{z}) \rangle = \varepsilon_{\alpha 
\beta} (z - \bar{z})^{- \frac{1}{2}},
\end{eqnarray}
\begin{eqnarray}
\langle S^{\dot{\alpha}} (y_1) S^{\dot{\beta}} (y_2) \rangle = 
\varepsilon^{\dot{\alpha} \dot{\beta}} (y_1 - y_2)^{-\frac{1}{2}},
\end{eqnarray}
\begin{eqnarray}
\langle  S^{\dot{\alpha}} (y_1) S^{\dot{\beta}} (y_2) 
S_{\alpha} (z) S_{\beta} (\bar{z}) \rangle = \varepsilon^{\dot{\alpha} 
\dot{\beta}} \varepsilon_{\alpha 
\beta} (y_1 - y_2)^{-\frac{1}{2}} (z - \bar{z})^{- \frac{1}{2}}.
\end{eqnarray}
If there are world sheet fermion inside the correlator, it should be 
carefully computed by evaluating the cocycle factor. Then, we find
\begin{eqnarray}
\langle S^{\dot{\alpha}} (y_1) \psi^{\mu} (y_2) S_{\alpha} (y_3) \rangle
= - \frac{1}{\sqrt{2}} (\bar{\sigma}^{\mu})^{\dot{\alpha} \beta} 
\varepsilon_{\beta \alpha} (y_1 - y_2)^{-\frac{1}{2}} (y_2 - y_3)^{- \frac{1}{2}}.
\end{eqnarray} 
The internal part is also computed to be
\begin{eqnarray}
\langle S_i (z) S_j (\bar{z}) \rangle = \varepsilon_{ij} (z - \bar{z})^{-\frac{1}{2}},
\end{eqnarray}
\begin{eqnarray}
\langle S^i (y_1) S^j (y_2) S^k (z) S^l (\bar{z}) \rangle 
&=& \left[ (y_1 - y_2) (y_1 - z) (y_1 - \bar{z}) (y_2 - z) (y_2 - \bar{z}) 
 (z - \bar{z}) \right]^{-\frac{1}{2}} \nonumber \\
& & \qquad \times \left[ \varepsilon^{il} \varepsilon^{jk} 
(y_1 - z) (y_2 - \bar{z}) - \varepsilon^{ik} \varepsilon^{jl} (y_2 - z) (y_1 
- \bar{z}) \right] \nonumber \\
&=& \left[ (y_1 - y_2) (y_1 - z) (y_1 - \bar{z}) (y_2 - z) (y_2 - \bar{z}) 
 (z - \bar{z}) \right]^{-\frac{1}{2}} \nonumber \\
& & \qquad \times \left[ - \varepsilon^{ij} \varepsilon^{kl} 
(y_1 - \bar{z}) (y_2 - z) + \varepsilon^{il} \varepsilon^{jk} (y_1 - y_2) 
(z - \bar{z}) \right]. \nonumber \\
\end{eqnarray}
If there are Lorentz generators in the correlator, one first should 
reduce it to the one which does not contain any Lorentz generator by the 
formula presented in \cite{KoLeLeSaWa}:
\begin{eqnarray}
& & \left\langle O^{(p)} (z_p) \dots O^{(j+1)} (z_{j+1}) :\psi^{M} 
\psi^{N} (z): O^{(j)} (z_j) \cdots O^{(1)} (z_1) \right\rangle \nonumber \\
& & \qquad = \sum_l \left\{ 
(M^{MN})^l_{\ l'} (z - z_l)^{-1} + ({M'}^{MN})^l_{\ l'} (z - z_l)^{-2} 
\right\} \nonumber \\
& & \qquad \quad  \times \left\langle O^{(p)} (z_p) \dots O^{(j+1)} (z_{j+1})  
 O^{(l')} (z_l) \cdots O^{(1)} (z_1) \right\rangle.
\end{eqnarray}
Here, $M,N$ is the ten-dimensional space-time indices and the matrices ${(M^{MN})}^l_{\ 
l'}$ and $({M'}^{MN})^l_{\ l'}$ are specified by the OPE
\begin{eqnarray}
:\psi^{M} \psi^{N} (z) : O^{(l)} (w) \sim \left[
(z - w)^{-2} ({M'}^{MN})^l_{\ l'} + (z - w)^{-1} (M^{MN})^l_{\ l'}
\right] O^{(l')} (w).
\end{eqnarray}
The space-time Lorentz generator $:\psi^{\mu} \psi^{\nu}:$ correlates only with the four-dimensional part, so we can find
\begin{eqnarray}
\langle S^{\dot{\alpha}} (z_1) \psi^{\mu} \psi^{\nu} (z_2) 
 S^{\dot{\beta}} (z_3) \rangle
%&=& - \frac{1}{2} \frac{1}{z_2 - z_1} ( \bar{\sigma}^{\mu 
% \nu})^{\dot{\alpha}}_{\ \dot{\gamma}} \langle S^{\dot{\gamma}} (z_1) 
% S^{\dot{\beta}} (z_3) \rangle  - \frac{1}{2} (\bar{\sigma}^{\mu \nu})^{\dot{\beta}}_{\ 
% \dot{\gamma}} \frac{1}{z_2 - z_3} \langle S^{\dot{\alpha}} (z_1) S^{\dot{\gamma}} (z_3) 
% \rangle \nonumber \\
= - \frac{1}{2} (\bar{\sigma}^{\mu \nu})^{\dot{\alpha} \dot{\beta}} 
 (z_1 -  z_3)^{\frac{1}{2}} (z_1 - z_2)^{-1} (z_2 - z_3)^{-1}.
\end{eqnarray} 
If there are two Lorentz generators in the correlator, we see
\begin{eqnarray}
& & \langle \psi^{\mu} \psi^{\nu} (y_1) \psi^{\rho} \psi^{\sigma} (y_2) 
 S_{\alpha} (z) S_{\beta} (\bar{z}) \rangle \nonumber \\
&=& - \varepsilon_{\alpha \beta} \left( \delta^{\mu \rho} \delta^{\nu 
			       \sigma} - \delta^{\mu \sigma} \delta^{\nu 
			       \rho}  \right) (y_1 - y_2)^{-2} (z - 
\bar{z})^{- \frac{1}{2}} \nonumber \\
& & - \frac{1}{4} (\sigma^{\mu \nu})_{\alpha}^{\ \gamma} ( \sigma^{\rho 
 \sigma})_{\gamma \beta} ( y_1 - z )^{-1} (y_2 - z)^{-1} (y_2 - \bar{z})^{-1} (z - 
 \bar{z})^{  \frac{1}{2}} \nonumber \\
& & - \frac{1}{4} ( \sigma^{\mu \nu} )_{\beta}^{\ \gamma} ( \sigma^{\rho 
 \sigma})_{\gamma \alpha} ( y_1 - \bar{z})^{-1} (y_2 - z)^{-1} (y_2 - 
 \bar{z})^{-1} ( z - \bar{z})^{\frac{1}{2}}.
\end{eqnarray}
In the same way, we find
\begin{eqnarray}
& & \langle \psi^{\mu} \psi^{\nu} (y_1) S^{\dot{\alpha}} (y_2) S^{\dot{\beta}} 
 (y_3) S_{\alpha} (z) S_{\beta} (\bar{z}) \rangle \nonumber \\
& & = \frac{1}{2} (y_2 - y_3)^{- \frac{1}{2}} (z - \bar{z})^{- \frac{1}{2}}
\left[ (\bar{\sigma}^{\mu \nu})^{\dot{\alpha} \dot{\beta}} 
 \varepsilon_{\alpha \beta} \frac{(y_2 - y_3)}{(y_1 - y_2) (y_1 - y_3)}
+ (\sigma^{\mu \nu})_{\alpha \beta} \varepsilon^{\dot{\alpha} \dot{\beta}}
\frac{(z - \bar{z})}{(y_1 - z) (y_1 - \bar{z})} \right]. \nonumber \\
\end{eqnarray}
In the case of the internal ''Lorentz generator'' $:\Psi 
\overline{\Psi}:$, the same formula can be used. The result is
\begin{eqnarray}
 \langle : \Psi \overline{\Psi} (y_1) : \overline{\Psi} (y_2) 
\Psi (y_3) \rangle = \frac{1}{(y_1 - y_2) (y_1 - y_3)}.
\end{eqnarray} 
%\begin{eqnarray}
%\langle : \Psi \overline{\Psi} (y_1) : \overline{\Psi} (y_2) S^{(-)} (z) 
% S^{(-)} (\bar{z}) \rangle
%&=& \frac{1}{y_1 - y_2} \langle e^{\phi_3 (y_2)} e^{- \frac{1}{2} \phi_3 (z)} e^{- \frac{1}{2} \phi_3 (\bar{z})}
%\rangle \nonumber \\
%& & - \frac{1}{2} \frac{1}{y_1 - z} \langle e^{\phi_3 (y_2)} 
%e^{- \frac{1}{2} \phi_3 (z)} e^{- \frac{1}{2} \phi_3 (\bar{z})} \rangle 
%\nonumber \\
%& & - \frac{1}{2} \frac{1}{y_1 - \bar{z}} \langle e^{\phi_3 (y_2)} 
%e^{- \frac{1}{2} \phi_3 (z)} e^{- \frac{1}{2} \phi_3 (\bar{z})} \rangle 
%\nonumber \\
%&=& \frac{1}{2} (y_2 - z)^{-\frac{1}{2}} (y_2 - \bar{z})^{-\frac{1}{2}}
%(z - \bar{z})^{\frac{1}{4}} \left[ \frac{2}{y_1 - y_2} - \frac{1}{y_1 - 
%z} - \frac{1}{y_1 - \bar{z}}   \right]. \nonumber \\
%\end{eqnarray}
%\begin{eqnarray}
%& & \langle : \Psi \overline{\Psi} (y_1) : : \Psi \overline{\Psi} (y_2) : \overline{\Psi} (y_2) S^{(-)} (z) 
% S^{(-)} (\bar{z}) \rangle \nonumber \\
%& & = (y_1 - y_2)^{-2} (y_3 - z)^{-\frac{1}{2}}
%(y_3 - \bar{z})^{-\frac{1}{2}} (z - \bar{z})^{\frac{1}{4}} \nonumber \\
%& & + \frac{1}{4} (y_3 - z)^{- \frac{1}{2}} (y_3 - \bar{z})^{-\frac{1}{2}}
%(z - \bar{z})^{\frac{1}{4}}  \left[ \frac{2}{y_2 - y_3} - \frac{1}{y_2-
% z} - \frac{1}{y_2 - \bar{z}}  \right]  \left[ \frac{2}{y_1 - y_3} - \frac{1}{y_1-
% z} - \frac{1}{y_1 - \bar{z}}  \right] . \nonumber \\
%\end{eqnarray}
{}From the ten-dimensional calculation, we find
\begin{eqnarray}
& & \langle S_{\gamma} (y_1) S^{\dot{\delta}} (y_2) \psi^{\mu} (y_3) 
S_{\alpha} (z) S_{\beta} (\bar{z}) \rangle \nonumber \\
& & = \frac{1}{\sqrt{2}} (y_1 - z)^{- \frac{1}{2}} (y_1 - \bar{z})^{- \frac{1}{2}}
(y_2 - y_3)^{-\frac{1}{2}} (y_1 - y_3)^{\frac{1}{2}} (y_3 - z)^{\frac{1}{2}} (y_3 - \bar{z})^{\frac{1}{2}} 
(z - \bar{z})^{- \frac{1}{2}} \nonumber \\
& & \qquad \times \left[
(\sigma^{\mu})_{\gamma}^{\ \dot{\delta}} \varepsilon_{\alpha \beta} 
(y_1 - y_3)^{-1} + (\sigma^{\mu})_{\alpha}^{\ \dot{\delta}} 
\varepsilon_{\gamma \beta} (y_3 - z)^{-1} - (\sigma^{\mu})_{\beta}^{\ \dot{\delta}}
\varepsilon_{\gamma \beta} (y_3 - \bar{z})^{-1} 
\right].
\end{eqnarray}
For $X^{\mu}$ fields, the general formula 
\cite{KoLeLeSaWa} is useful:
\begin{eqnarray}
\left\langle \prod_{\{l\}} \varepsilon^{(l)} \cdot 
\partial X (z_l) \prod_{j} e^{i \sqrt{2 \pi \alpha'} p^{(j)} \cdot X 
(z_j)} \right\rangle 
= \prod_{\{ l \}} \left[ \left. \frac{\partial}{\partial z_l} 
   \right|_{ \sqrt{2 \pi \alpha'} p^{(l)} \to \varepsilon^{(l)} }  \right]
\prod_{i>j} (z_i - z_j)^{2 \pi \alpha' p^{(i)} \cdot p^{(j)}}.
\end{eqnarray}
By using this formula, we find
\begin{eqnarray}
& & \left\langle A_{\mu} (p_1) \partial X^{\mu} (y_1) \prod_{j=1}^n e^{i \sqrt{2 \pi \alpha'} p_j \cdot X (y_j)} 
\right\rangle 
= \prod_{i<j} (y_i - y_j)^{ 2 \pi \alpha' p_i \cdot p_j} \times i (2 \pi \alpha')^{\frac{1}{2}}
\sum_{j=2}^n \left[ \frac{A_{\mu} (p_1) p^{\mu}_j}{y_1 - y_j} \right],\label{partial}
\nonumber \\
\end{eqnarray}
and
\begin{eqnarray}
& & \left\langle A_{\mu} (p_1) \partial X^{\mu} (y_1) e^{i \sqrt{2 \pi 
 \alpha'} p_1 \cdot X (y_1)} A_{\nu} (p_2) \partial X^{\nu} (y_2) e^{i \sqrt{2 \pi 
 \alpha'} p_2 \cdot X (y_2)} e^{i \sqrt{2 \pi \alpha'} p_3 \cdot X 
 (y_3)} \right\rangle \nonumber \\
& & \qquad =  (y_1 - y_2)^{ 2 \pi \alpha' p_1 \cdot p_2} (y_1 - 
 y_3)^{ 2 \pi \alpha' p_1 \cdot p_3} (y_2 - y_3)^{ 2 \pi \alpha' p_2 
 \cdot p_3} \nonumber \\
 & & \qquad \times \left[ \frac{A_{\mu} (p_1) A^{\mu} (p_2)}{(y_1 - 
 y_2)^2} + \frac{(2 \pi \alpha') A_{\mu} (p_1) p_2^{\mu} A_{\nu} (p_2) p_1^{\nu}}{(y_1 - y_2)^2}
+ \frac{(2 \pi \alpha') A_{\mu} (p_1) p^{\mu}_3 
A_{\nu} (p_2) p_1^{\nu} }{(y_1 - y_2) (y_1 - y_3)} 
\right. 
\nonumber \\
& & \qquad \qquad \qquad  \left.  - \frac{(2 \pi 
\alpha') A_{\mu} (p_1) p^{\mu}_2 A_{\nu} (p_2) p^{\nu}_3}{(y_1 - y_2) 
(y_2 - y_3)} - \frac{(2 \pi \alpha') A_{\mu} (p_1) p^{\mu}_3 A_{\nu} 
(p_2) p^{\nu}_3}{(y_1 - y_3)(y_2 - y_3)} \right] .
\end{eqnarray}

\end{appendix}


\begin{thebibliography}{99}

\bibitem{OoVa} 
H.~Ooguri and C.~Vafa,
%``The C-deformation of gluino and non-planar diagrams'', 
Adv. Theor. Math. Phys. {\bf 7} (2003) 53,
hep-th/0302109; 
%``Gravity induced C-deformation,''
Adv. Theor. Math. Phys. {\bf 7} (2004) 405, hep-th/0303063. 

\bibitem{BeSe}
N.~Berkovits and N.~Seiberg,
%``Superstrings in graviphoton background and N = 1/2 + 3/2
%        supersymmetry'',
JHEP {\bf 0307} (2003) 010,
hep-th/0306226. 

\bibitem{DeGrNi} J.~de Boer, P.~A.~Grassi and P.~van Nieuwenhuizen,
%``Non-commutative superspace from string theory'',
Phys. Lett. {\bf B574} (2003) 98, hep-th/0302078. 

\bibitem{Se} 
N.~Seiberg,
%``Noncommutative superspace, $N=1/2$ supersymmetry, field theory and
%string theory'',
JHEP {\bf 0306}, 010 (2003),
hep-th/0305248. 

\bibitem{ScNi}
J.~H.~Schwarz and P.~Van Nieuwenhuizen,
%``Speculations Concerning A Fermionic Substructure Of Space-Time,''
Lett.\ Nuovo Cim.\  {\bf 34}, 21 (1982). 

\bibitem{KlPeTa}
D.~Klemm, S.~Penati and L.~Tamassia,
%``Non(anti)commutative superspace,''
Class.\ Quant.\ Grav.\  {\bf 20} (2003) 2905,
hep-th/0104190; \\ 
S.~Ferrara and M.~A.~Lledo,
%``Some aspects of deformations of supersymmetric field theories,''
JHEP {\bf 0005} (2000) 008,
hep-th/0002084; \\
S.~Ferrara, M.~A.~Lledo and O.~Macia,
%``Supersymmetry in noncommutative superspaces'', 
JHEP {\bf 0309} (2003) 068,
hep-th/0307039.

\bibitem{Billo-1/2}
M.~Bill\'o, M.~Frau, I.~Pesando and A.~Lerda,
%`` $N=1/2$ Gauge theory and its instanton moduli space from open strings
%in RR background'',
JHEP {\bf 0405} (2004) 023,
hep-th/0402160.

\bibitem{IvLeZu}
E. Ivanov, O. Lechtenfeld and B. Zupnik,
%``Nilpotent Deformations of $N=2$ Superspace''
JHEP {\bf 0402} (2004) 012,
hep-th/0308012;
%``Non-anticommutative deformation of N = (1,1) hypermultiplets,''
Nucl.\ Phys.\ B {\bf 707} (2005) 69, 
hep-th/0408146. 

\bibitem{FeSo}
S.~Ferrara and E.~Sokatchev, 
%``Non-anticommutative $N=2$ super-Yang-Mills theory with singlet
%deformation'',
Phys. Lett. {\bf B579} (2004) 226, hep-th/0308021.

\bibitem{FeIvLeSoZu}
S.~Ferrara, E.~Ivanov, O.~Lechtenfeld, E.~Sokatchev and B.~Zupnik,
%``Non-anticommutative chiral singlet deformation of N = (1,1) gauge 
%theory'', 
Nucl.\ Phys.\ B {\bf 704} (2005) 154, hep-th/0405049.

\bibitem{Castro}
A.~De~Castro, E.~Ivanov, O.~Lechtenfeld, L.~Quevedo,
% Non-singlet Q-deformation of the N=(1,1) gauge multiplet in 
% harmonic superspace.
Nucl. Phys. {\bf B747} (2006) 1, hep-th/0510013;\\
A.~De~Castro and L.~Quevedo
%''Non-singlet Q-deformed N=(1,0) and N=(1,1/2) U(1) actions''
Phys. Lett. {\bf B639} (2006) 117, hep-th/0605187. 

\bibitem{ArItOh2}
T.~Araki, K.~Ito and A.~Ohtsuka, 
%``$\mathcal{N}=2$ Supersymmetric $U(1)$ gauge theories in noncommutative 
%harmonic superspace'',
JHEP {\bf 0401} (2004) 046, hep-th/0401012; 
Phys. Lett. {\bf B606} (2005) 202, hep-th/0410203.

\bibitem{ArIt3}
T.~Araki and K.~Ito, 
%``Singlet Deformation and Non(anti)commutative N=2 Supersymmetric U(1)
%       Gauge Theory'', 
Phys. Lett. {\bf B595} (2004) 513, hep-th/0404250.

\bibitem{Billo-N=2}
M.~Bill\'o, M.~Frau, F.~Fucito and A.~Lerda,
%''Instanton calculus in R-R background and the topological string''
hep-th/0606013.

\bibitem{ArItOh1}
T.~Araki, K.~Ito and A.~Ohtsuka, 
%``Supersymmetric gauge theories on noncommutative superspace'',
Phys. Lett. {\bf B573} (2003) 209, 
hep-th/0307076. 

\bibitem{Ne}
N.A.~Nekrasov, Adv. Theor. Math. Phys. {\bf 7} (2004) 831, hep-th/0206161.

\bibitem{KoLeLeSaWa}
V.~A.~Kostelecky, O.~Lechtenfeld, W.~Lerche, S.~Samuel and S.~Watamura,
%''Conformal Techniques, Bosonization And Tree Level String Amplitudes'',
Nucl. Phys. {\bf B288} (1987) 173. 

\bibitem{FMS}
D.~Friedan, E.~J.~Martinec and S.~H.~Shenker,
%''Conformal Invariance, Supersymmetry And String Theory'',
Nucl. Phys. {\bf B271} (1986) 93. 

\bibitem{WeBa}
J.~Wess and J.~Bagger, ``Supersymmetry and
Supergravity,''
Princeton University Press, 1992.

\bibitem{VMLRM}
P.~Di~Vecchia, L.~Magnea, A.~Lerda, R.~Russo and R.~Marotta, 
%''String Techniques for the Calculation of Renormalization 
%Constants in Field Theory'', 
Nucl. Phys. {\bf B469} (1996) 235, hep-th/9601143.

\bibitem{DIS}
M.~Bill\'o, M.~Frau, I.~Pesando, F.~Fucito, A.~Lerda and A.~Liccardo,
JHEP {\bf 0302} (2003) 045, hep-th/0211250

\bibitem{ADS}
J.~J.~Atick, L.~J.~Dixon and A.~Sen,
%''String Calculation of Fayet-Iliopoulos D-terms in Arbitrary 
%	Supersymmetric Compactifications'', 
Nucl. Phys. {\bf B292} (1987) 109.\\
M.~Dine, I.~Ichinose and N.~Seiberg, 
%''F Terms and D Terms in String Theory''
Nucl. Phys. {\bf B293} (1987) 253.

\bibitem{Billo-quiver}
M.~Bill\'o, M.~Frau, F.~Lonegro and A.~Lerda,
%`` N=1/2 quiver gauge theories from open strings with R-R fluxes''
JHEP {\bf 0505} (2005) 047, hep-th/0502084.

\bibitem{ArItOh5}
T.~Araki, K.~Ito and A.~Ohtsuka, 
%``Non(anti)commutative $\mathcal{N}=(1,1/2)$ supersymmetric $U(1)$ gauge
% theory'',
JHEP {\bf 0505} (2005) 074, hep-th/0503224.

\bibitem{Im1}
A.~Imaanpur and S.~Parvizi,
%''N = 1/2 super Yang-Mills theory on Euclidean AdS(2) x S**2'',
JHEP {\bf 0407} (2004) 010, hep-th/0403174.

\bibitem{Im2}
A.~Imaanpur, 
%``Supersymmetric D3-branes in five-form flux'',
JHEP {\bf 0503} (2005) 030, hep-th/0501167;\\
R.~Abbaspur and A.~Imaanpur,
%``Nonanticommutative deformation of N=4 SYM theory: The Myers effect and vacuum states.'',
JHEP {\bf 0601} (2006) 017, hep-th/0509220. 

%\bibitem{ItKoSa}
%K.~Ito, Y.~Kobayashi and S.~Sasaki, work in progress.



\end{thebibliography}
\end{document}